\documentclass[twocolumn,twocolappendix]{aastex631}

\newcommand{\BOne}{Full Limb}
\newcommand{\BTwo}{Jet}

\usepackage[version=4]{mhchem}
\let\tablenum\relax
\usepackage{siunitx}
\usepackage{color}
\usepackage{gensymb}

\newcommand{\espalias}{E24}
\defcitealias{espinoza2024inhomogeneous}{\espalias}

\begin{document}

\title{Limb Asymmetries on WASP-39b: A Multi-GCM Comparison of Chemistry, Clouds, and Hazes}

\author[0000-0001-8342-1895]{Maria E. Steinrueck}
\altaffiliation{51 Pegasi b fellow}
\affiliation{Department of Astronomy \& Astrophysics, University of Chicago, Chicago, IL 60637, USA}
\affiliation{Max Planck Institute for Astronomy, 69117 Heidelberg, Germany}

\author[0000-0002-2454-768X]{Arjun B. Savel}
\affiliation{Department of Astronomy, University of Maryland, College Park, MD 20742, USA}

\author[0000-0002-4997-0847]{Duncan A. Christie}
\affiliation{Max Planck Institute for Astronomy, 69117 Heidelberg, Germany}

\author[0000-0001-9355-3752]{Ludmila Carone}
\affiliation{Space Research Institute, Austrian Academy of Sciences, Schmiedlstrasse 6, 8042 Graz, Austria}

\author[0000-0002-8163-4608]{Shang-Min Tsai}
\affiliation{Department of Earth and Planetary Sciences, University of California, Riverside, CA, USA}
\affiliation{Institute of Astronomy \& Astrophysics, Academia Sinica, Taipei 10617, Taiwan}

\author[0000-0001-9499-2183]{Can Ak{\i}n}
\affiliation{Center for Space and Habitability, University of Bern, Gesellschaftsstrasse 6, 3012 Bern, Switzerland}
\affiliation{Universitäts-Sternwarte München, Fakultät für Physik der Ludwig-Maximilians-Universität, Scheinerstraße 1, 81679 München, Deutschland}

\author[0000-0002-2984-3250]{Thomas D. Kennedy}
\affiliation{Department of Astronomy and Astrophysics, University of Michigan, Ann Arbor, MI, 48109, USA}

\author[0000-0003-1285-3433]{Sven Kiefer}
\affiliation{Institute of Astronomy, KU Leuven, Celestijnenlaan 200D, 3001 Leuven, Belgium}
\affiliation{Space Research Institute, Austrian Academy of Sciences, Schmiedlstrasse 6, 8042 Graz, Austria}
\affiliation{Institute for Theoretical Physics and Computational Physics, Graz University of Technology, Petersgasse 16, 8010 Graz, Austria}
\affiliation{University of Texas at Austin, Department of Astronomy, 2515 Speedway C1400, Austin, TX 78712, USA}

\author[0000-0002-1871-2773]{David A. Lewis}
\affiliation{Centre for Science at Extreme Conditions and Scottish Universities Physics Alliance, School of Physics and Astronomy, The University of Edinburgh, Edinburgh, UK}
\affiliation{School of GeoSciences, The University of Edinburgh, Edinburgh, UK}

\author[0000-0003-3963-9672]{Emily Rauscher}
\affiliation{Department of Astronomy and Astrophysics, University of Michigan, Ann Arbor, MI, 48109, USA}

\author[0000-0002-8956-2047]{Dominic Samra}
\affiliation{Department of Astronomy \& Astrophysics, University of Chicago, Chicago, IL 60637, USA}

\author[0000-0002-9705-0535]{Maria Zamyatina}
\affiliation{Department of Physics and Astronomy, Faculty of Environment, Science and Economy, University of Exeter, Exeter EX4 4QL, UK}

\author[0000-0002-3034-8505]{Anastasia Triantafillides}
\affiliation{Department of Astronomy, University of Maryland, College Park, MD 20742, USA}
\affiliation{Department of Astronomy, University of Wisconsin–Madison, 475 N. Charter St., Madison, WI 538706, USA}

\author[0000-0001-7578-969X]{Robin Baeyens}
\affiliation{Anton Pannekoek Institute for Astronomy, University of Amsterdam, Science Park 904, 1098 XH, Amsterdam, The Netherlands}

\author[0000-0002-1397-8169]{Leonardos Gkouvelis}
\affiliation{Instituto de Astrofísica de Andalucía (IAA-CSIC), 
Glorieta de la Astronomía s/n, E-18008 Granada, Spain}

\author[0009-0009-7667-7003]{David Haegele}
\affiliation{Max Planck Institute for Astronomy, 69117 Heidelberg, Germany}

\author[0000-0002-8275-1371]{Christiane Helling}
\affiliation{Space Research Institute, Austrian Academy of Sciences, Schmiedlstrasse 6, 8042 Graz, Austria}
\affiliation{Institute for Theoretical Physics and Computational Physics, Graz University of Technology, Petersgasse 16, 8010 Graz, Austria}

\author[0000-0001-6707-4563]{Nathan J. Mayne}
\affiliation{Department of Physics and Astronomy, Faculty of Environment, Science and Economy, University of Exeter, Exeter EX4 4QL, UK}

\author[0000-0002-4250-0957]{Diana Powell}
\affiliation{Department of Astronomy \& Astrophysics, University of Chicago, Chicago, IL 60637, USA}

\author[0000-0001-8206-2165]{Michael T. Roman}
 \affiliation{Facultad de Ingeniería y Ciencias, Universidad Adolfo Ibáñez, Santiago, Chile 8170121}
 \affiliation{School of Physics and Astronomy, University of Leicester, Leicester, United Kingdom LE1 7RH}

\author[0000-0002-6980-052X]{Hayley Beltz}
\affiliation{Department of Astronomy, University of Maryland, College Park, MD 20742, USA}

\author[0000-0001-9513-1449]{Nestor Espinoza}
\affiliation{Space Telescope Science Institute, 3700 San Martin Drive, Baltimore, MD 21218, USA}
\affiliation{William H. Miller III Department of Physics and Astronomy, Johns Hopkins University, Baltimore, MD 21218, USA}

\author[0000-0003-1907-5910]{Kevin Heng}
\affiliation{Universitäts-Sternwarte München, Fakultät für Physik der Ludwig-Maximilians-Universität, Scheinerstraße 1, 81679 München, Deutschland}
\affiliation{University College London, Department of Physics \& Astronomy, Gower St, London, WC1E 6BT, United Kingdom}
\affiliation{Astronomy \& Astrophysics Group, Department of Physics, University of Warwick, Coventry CV4 7AL, United Kingdom}

\author[0000-0003-2329-418X]{Nicolas Iro}
\affiliation{Institute of Planetary Research, German Aerospace Center (DLR), Rutherfordstrasse 2, D-12489 Berlin, Germany}

\author[0000-0002-1337-9051]{Eliza M.-R. Kempton}
\affiliation{Department of Astronomy \& Astrophysics, University of Chicago, Chicago, IL 60637, USA}
\affiliation{Department of Astronomy, University of Maryland, College Park, MD 20742, USA}

\author[0000-0003-0514-1147]{Laura Kreidberg}
\affiliation{Max Planck Institute for Astronomy, 69117 Heidelberg, Germany}

\author[0000-0002-4207-6615]{James Kirk}
\affiliation{Department of Physics, Imperial College London, Prince Consort Road, SW7 2AZ, London, UK}

\author[0000-0002-8517-8857]{Matthew M. Murphy}
\affiliation{Steward Observatory, 933 North Cherry Avenue, Tucson, AZ 85721, USA}

\author[0000-0002-3627-1676]{Benjamin V.\ Rackham}
\affiliation{Department of Earth, Atmospheric and Planetary Sciences, Massachusetts Institute of Technology, 77 Massachusetts Avenue, Cambridge, MA 02139, USA}
\affiliation{Kavli Institute for Astrophysics and Space Research, Massachusetts Institute of Technology, Cambridge, MA 02139, USA}

\author[0000-0003-2278-6932]{Xianyu Tan}
\affiliation{Tsung-Dao Lee Institute, Shanghai Jiao Tong University, 1 Lisuo Road, Shanghai 200127, People’s Republic of China}

\begin{abstract}

With JWST, observing separate spectra of the morning and evening limbs of hot Jupiters has finally become a reality. The first such observation was reported for WASP-39b, where the evening terminator was observed to have a larger transit radius by about 400~ppm and a stronger 4.3\,$\mu$m \ce{CO2} feature than the morning terminator. Multiple factors, including temperature differences, photo/thermochemistry, clouds and hazes, could cause such limb asymmetries. To interpret these new limb asymmetry observations, a detailed understanding of how the relevant processes affect morning and evening spectra grounded in forward models is needed. Focusing on WASP-39b, we compare simulations from five different general circulation models (GCMs), including one simulating disequilibrium thermochemistry and one with cloud radiative feedback, to the recent WASP-39b limb asymmetry observations. We also post-process the temperature structures of all simulations with a 2D photochemical model and one simulation with a cloud microphysics model. Although the temperatures predicted by the different models vary considerably, the models are remarkably consistent in their predicted morning--evening temperature differences. Several equilibrium-chemistry simulations predict strong methane features in the morning spectrum, not seen in the observations. When including disequilibrium processes, horizontal transport homogenizes methane, and these methane features disappear. However, even after including photochemistry and clouds, our models still cannot reproduce the observed ${\sim}2000$\,ppm asymmetry in the \ce{CO2} feature. A combination of factors, such as varying metallicity and unexplored parameters in cloud models, may explain the discrepancy, emphasizing the need for future models integrating cloud microphysics and feedback across a broader parameter space.

\end{abstract}


\section{Introduction} \label{sec:intro}
Transit spectroscopy probes the day--night boundary (terminator region) of the atmospheres of exoplanets. When interpreting transmission spectra, it is typically assumed that this region is homogeneous, with properties that depend only on the radial coordinate. However, for tidally synchronized extrasolar giant planets, which have been the focus of most observational campaigns aimed at characterizing exoplanet atmospheres to date, the terminator region is expected to be inhomogeneous: atmospheric dynamics predicts temperature differences between the evening terminator (trailing limb in transit) and morning terminator (leading limb in transit) for these planets \citep[e.g.,][]{FortneyEtAl2010,DobbsDixonEtAl2012,2019A&A...631A..79H, caldas2019, pluriel2022}. In particular, most hot Jupiters are expected to feature a super-rotating equatorial jet \citep{ShowmanGuillot2002,ShowmanPolvani2011, TsaiEtAl2014} transporting hot air eastward from the day side of the planet towards the evening terminator. At the morning terminator, in contrast, the jet carries cold air flowing from the night side towards the day side. In addition, in many 3D simulations of hot Jupiters, large mid-latitude gyres form east of the antistellar point, trapping cold air. Often, these gyres lead to even lower temperatures at the morning terminator at mid-latitudes compared to the equatorial region \citep[e.g.,][]{ShowmanEtAl2009}.

The differences in temperature of the two limbs lead to differences in the chemical processes, leading to predictions of differences in gas-phase abundances \citep[e.g.,][]{FortneyEtAl2010,MosesEtAl2011,KatariaEtAl2016} as well as differences in cloudiness \citep[e.g.,][]{LineParmentier2016,VonParisEtAl2016,LeeEtAl2016,KemptonEtAl2017,PowellEtAl2019TransitSignatures,Helling2020_W43b,Roman2021,Carone2023}. In addition, atmospheric transport strongly affects the abundances of chemical species and aerosols and can either drive limb asymmetries, for example by concentrating small photochemical hazes near the morning terminator \citep{SteinrueckEtAl2021}, or homogenize the abundances of chemical species with long reaction timescales, for example methane \citep[e.g.,][]{CooperShowman2006,AgundezEtAl2014,Drummond2018a, mendonca_2018b, baeyens2021, ZamyatinaEtAl2024WASP-96b} or SO$_2$ \citep{Tsai2023b}.
Directly detecting differences between the morning and evening terminator thus presents an opportunity to test our understanding of atmospheric dynamics of hot Jupiters, to quantify biases that arise when using 1D models in interpreting transmission spectra \citep[e.g.,][]{LineParmentier2016,fu2025}, and to obtain additional observational constraints for cloud and photochemical models. 

\subsection{Limb Asymmetry Measurements}
Previously, differences between the morning and evening terminator have been detected for multiple ultra-hot Jupiters through high-resolution cross-correlation observations  \citep{EhrenreichEtAl2020,PrinothEtAlWASP-189b2023,KesseliEtAl2022,PelletierEtAl2023WASP-76b, SeidelEtAl2025}.
These detections took advantage of the fact that ultra-hot Jupiters rotate substantially over the course of the transit and observed a shift in the spectrum as the morning dayside rotated out of view and the evening dayside rotated into view. These detections have been ground-breaking for studying spatial differences in exoplanets. However, this technique also has limitations: so far, it has mostly been limited to ultra-hot Jupiters orbiting bright stars because of the high signal-to-noise ratio required for ground-based high-resolution observations and because of the geometric constraints (a large degree of rotation during the transit is required) \citep{Wardenier2022}. In addition, spectral information on the continuum is lost during the cross-correlation process, and the resulting observational signature can be quite complex to interpret \citep[e.g.,][]{Savel2022, Savel2023, Wardenier2021, Wardenier2023,Beltz2023}.

A promising complementary approach is transit mapping techniques, which attempt to recover information separately from the morning and evening limbs via transit photometry or spectroscopy and primarily utilize information from the ingress and egress. This approach has long been suggested as a way to observe thermal, chemical and cloud limb asymmetries
\citep{FortneyEtAl2010,DobbsDixonEtAl2012,KemptonEtAl2017,PowellEtAl2019TransitSignatures}. 
While it has been used in high-resolution cross-correlation observations before \citep{LoudenWheatley2015}, JWST for the first time opens up the possibility to directly observe limb differences in low-resolution spectra, preserving the spectral shape and thus information on temperature structure and abundances. In one of the first exoplanet observations with JWST, \citet{RustamkulovEtAl2023} reported a detection of a wavelength-dependent shift of the mid-transit time in the NIRSpec PRISM transit observation of hot Jupiter WASP-39b as an indication of limb asymmetries, as predicted by \citet{DobbsDixonEtAl2012}. \citet[hereafter \espalias]{espinoza2024inhomogeneous}  analyzed this dataset for limb asymmetries using the \texttt{catwoman} software package \citep{JonesEspinoza2020catwoman} which approximates the planet as two half-circles rather than a single circle. They were able to derive separate spectra for the morning and evening sides of the planet. Their analysis showed an offset between the morning and evening spectra of $405 \pm 88$~ppm as well as differences in the spectral shape, most notably a much stronger CO$_2$ feature for the evening terminator. One-dimensional retrievals on the spectra suggested a temperature difference of $177^{+65}_{-57}$~K.

At the same time, \citet{MurphyEtAl2024WASP-107b} performed a similar analysis for JWST/NIRCAM F322W2 observations of the low-density hot Neptune WASP-107b, also finding differences in absolute transit depth and spectral shape between morning and evening terminators. These results were later extended to a broader wavelength range in \citet{murphy2025} and were interpreted as evidence for inhomogeneous cloud coverage. The latter study also reported tentative evidence for abundance variations in \ce{SO2} and \ce{CO2}.  More recently, \citet{TadaEtAl2025WASP-39bLimbAsymmetries} reported differences in the CO$_2$ feature and tentative differences in the SO$_2$ feature in a different set of observations of WASP-39b using NIRSpec/G395H, and \citet{MukherjeeEtAl2025WASP-94Ab} detected strong terminator differences on the hotter WASP-94Ab using NIRISS/SOSS. \citet{fu2025} examined the NIRISS/SOSS spectra of nine exoplanets for limb asymmetries and found statistically significant variations in three planets, (WASP-94Ab, WASP-39b and WASP-17b).

\subsection{The Need for 3D Modeling}
To interpret these exciting new observations, a comparison with the predictions of 3D general circulation models (GCMs) is indispensable. While retrievals can quantify differences in spectra and attribute them to differences in abundances, temperature, or cloudiness with some statistical likelihood, only multi-dimensional forward models that simulate the atmospheric dynamics like GCMs can provide physical insight into which processes are the primary drivers of these limb differences. However, such a comparison also comes with challenges: even a single GCM simulation is computationally expensive, making fitting infeasible, and currently, no single GCM is equipped to model all processes that are expected to affect limb asymmetries simultaneously. In addition, GCM predictions for the same planet, even with similar input parameters, can vary substantially depending on inherent model differences, including for example differences in the dynamical cores, numerical dissipation, treatment of radiative transfer and the inclusion of additional physics such as chemical kinetics or radiative feedback from clouds. 

\subsection{Goals of This Work}
While \citetalias{espinoza2024inhomogeneous} and \citet{MukherjeeEtAl2025WASP-94Ab} included brief comparisons to the results of GCM simulations, a detailed analysis of GCM simulations was beyond the scope of these studies. The aims of this work are: 1) to quantify the extent to which GCMs of hot Jupiters in their typical configurations vary in their predictions regarding limb asymmetries, 2)
to provide a comprehensive examination of how various processes predicted to influence limb asymmetries affect the simulated outcomes for the same planet from a forward-modeling perspective,
3) to compare this set of GCM simulations with recent observations of limb asymmetries. We choose to focus on WASP-39b, the first planet for which limb asymmetries in low-resolution spectra were reported and arguably the best-studied exoplanet with JWST.

To accomplish this goal of exploring the mechanisms driving limb asymmetry, we analyze simulations conducted with five different GCMs, described in Section \ref{sec:methods}. First, we compare the clear-atmosphere, equilibrium chemistry version of each GCM to assess the impact of temperature structure and the variations between different GCMs (Section \ref{sec:temperaturestructure}). We then examine the impact of disequilibrium chemistry on limb asymmetries using two methods: a GCM capable of simulating transport-induced disequilibrium chemistry and a 2D photochemical model using the thermal structure and winds from the simulations of all five GCMs included in this study as input (Section \ref{sec:disequilib}). Many different approaches exist to include condensate clouds in GCMs. We choose to use two complementary cloud models: a parametrized equilibrium condensation model to examine the impact of radiative feedback on the temperature structure within a GCM, and a post-processed microphysics model to better examine how cloud particle sizes and composition vary between morning and evening terminator. The results from both models are discussed in Section \ref{sec:clouds}. We then use a radiatively passive tracer-based model to simulate the 3D distribution of photochemical hazes (Section \ref{sec:hazes_results}). Finally, we also explore combining disequilibrium chemistry and clouds simultaneously in post-processing (Section \ref{sec:photochem+clouds}). In Section \ref{sec:discussion}, we discuss possible explanations for the remaining discrepancies between our models and the observations of \citetalias{espinoza2024inhomogeneous}, and avenues for future work, and in Section \ref{sec:conclusions}, we summarize our conclusions.

\section{Methods} \label{sec:methods}

The modeling of WASP-39b used here encompasses both three-dimensional GCMs and the post-processing of those models with more physically complex cloud and photochemistry schemes.   We outline the various approaches here. 

\subsection{General Circulation Models} \label{subsec:methods_gcms}

\begin{deluxetable*}{ccp{3cm}p{4.5cm}c}

\tablecaption{Overview of General Circulation Models used in this Work}
\label{Tbl:GCMs}

\tablehead{\colhead{Model} & \colhead{Radiative Transfer} & \colhead{Gas-phase Chemistry} & \colhead{Cloud/Haze Model\tablenotemark{a} } & \colhead{Aerosol Feedback} } 
\startdata
ExoRad/IWF Graz  & Correlated-k & Equilibrium\tablenotemark{b} & Post-processed microphysical condensate cloud model with mixed grains. (see Section \ref{subsec:IWF_cloud_description}) & No \\
RM-GCM & Picket-fence & -- & Temperature-based equilibrium condensate cloud model with parametrized cloud thickness (see Section \ref{subsec:RMGCM_cloud_description})    & Yes \\
SPARC/MITgcm & Correlated-k & Equilibrium  & Passive haze tracers with a constant particle size (see Section \ref{subsec:hazes_description})& No \\
THOR & Double-gray & -- & --  & -- \\
Unified Model & Correlated-k & Equilibrium or chemical kinetics & --& -- \\
\enddata

\tablenotetext{a}{We use the term ``clouds'' to describe particles condensing out of the gas phase in a reversible process and the term ``hazes''to describe particles forming from photochemical byproducts.}
\tablenotetext{b}{For the post-processing of the cloudy case, elemental depletion from the clouds was taken into account in the equilibrium chemistry calculation.}

\end{deluxetable*}

To capture different aspects of the atmosphere of WASP-39b, five different general circulation models, each with an established history of use in simulating exoplanets, were run to examine the 3D structure of WASP-39b. As each GCM has different capabilities, this ensemble will allow for the atmosphere to be better characterized.  The included GCMs are summarized in Table \ref{Tbl:GCMs}, and the configurations and setups of each of the GCMs are outlined in the following subsections. All GCMs used the planetary parameters listed in Table \ref{tab:W-39bparameters}. An exception to this is the ExoRad/IWF~Graz model, which used the set of parameters listed in Table A.1 in \citet{Carone2023} internally within the GCM and cloud microphysics model, though the stellar and planetary radii from Table \ref{tab:W-39bparameters} were used in post-processing to generate spectra to ensure better comparability with the other models. The differences between the two sets of parameters are comparable to the observational uncertainty for each parameter and the impact on simulation results is expected to be small compared to the differences between simulations arising e.g., from different radiative transfer methods. Early analyses of JWST spectra of WASP-39b suggested a metallicity close to 10$\times$~solar based on fits to forward models \citep{ERSteam2023co2,Ahrer2023,AldersonEtAl2023WASP-39bG395H,Feinstein2023,RustamkulovEtAl2023}. We therefore adopt a metallicity of 10$\times$~solar in all simulations.  We note also that due to our goal to compare different GCMs in the configuration in which they are typically used in their respective research groups, the contributed simulations do not all use the same intrinsic temperature $T_\mathrm{int}$ for the planet. This results in differences in the simulations \citep[see, e.g.,][]{KomacekEtAl2022} beyond the differences in the physics being investigated, although we expect these effects to be secondary compared to the impact of, for example, aerosol opacity.

\begin{deluxetable*}{lrlp{8cm}}

\tablecaption{WASP-39b Planetary Parameters}
\label{tab:W-39bparameters}
\tablenum{}

\tablehead{\colhead{Quantity} & \colhead{Value} & \colhead{Units} & \colhead{Source}  } 
\startdata
Planetary Radius &  8.94$\times10^{7}$ &  m &  \citet{ManciniEtAl2018} \\
Stellar Radius &  6.48$\times10^{8}$ &  m &  Gaia DR3 \\
Gravity &  4.26 &  m s$^{-2}$ &  \citet{ManciniEtAl2018} \\
Planetary Rotation Period &  4.055 &  d &  \citet{FischerEtAl2016} \\
Zero-albedo Equilibrium Temperature &  1166 &  K &  \citet{ManciniEtAl2018} \\
Metallicity &  10$\times$ solar &   &  \citet{ERSteam2023co2}  \\ 
 \\
\enddata

\end{deluxetable*}

\subsubsection{ExoRad}
\label{subsubsec:methods_gcm_ExpeRT/MITgcm}

The ExoRad GCM (also referred to as the expeRT/MITgcm)  uses the dynamical core of the MITgcm \citep{Adcroft2004}, solving the hydrostatic primitive equations on a C32\footnote{equivalent to 128 $\times$ 64 grid points in the longitude and latitude directions} cubed-sphere grid \citep{Adcroft2004}. It resolves the atmosphere at pressures less than 100~bar on 41 log-spaced cells between $1 \times 10^{-5}$ bar and $100$~bar. At pressures greater than 100~bar, six linearly spaced grid cells between $100$~bar and $700$~bar are added. The model uses a sponge layer that damps the horizontal velocity to its zonal mean between $p=10^{-4}-10^{-5}$~bar and also applies basal drag to pressure layers deeper than $500$~bar \citep{Carone2020,Schneider2022}. In addition, like in SPARC/MITgcm, a fourth-order Shapiro filter is applied.

The GCM is coupled to a non-gray radiative transfer scheme based on petitRADTRANS \citep{Molliere2019}. Fluxes are recalculated every 4th dynamical timestep. Stellar irradiation is taken from the PHOENIX model atmosphere suite \citep{allard1995, hauschildt1999, husser2013}. The GCM operates on a pre-calculated grid of correlated-k-binned opacities, where the opacities are taken from the ExoMol database \citep{Chubb2021} and precalculated offline on a grid of 1000 logarithmically spaced temperature points between 100 K and 4000 K for every vertical layer. We further include the same species as shown in \citet{Schneider2022} except TiO and VO to avoid the formation of a temperature inversion in the upper atmosphere. These are: \ce{H2O}, \ce{CH4}, \ce{CO2}, \ce{NH3}, \ce{CO}, \ce{H2S}, \ce{HCN}, \ce{PH3}, \ce{FeH}, Na, and K. For Rayleigh scattering, the opacities are \ce{H2} and \ce{He}, and we add the following CIA opacities: \ce{H2}--\ce{H2} and \ce{H2}--{He}. The sources for all the opacity data are found in Table \ref{tab:opacitydata} in Appendix \ref{sec:opacitydata}. For the radiative transfer calculations in the GCM, we use a similar wavelength resolution as SPARC/MITgcm \citep{KatariaEtAl2013}, but incorporate 16 instead of 8 k-coefficients \citep[S1 case in][]{Schneider2022}. For WASP-39b, $10\times$ solar metallicity and an interior heat flux corresponding to a temperature of $T_\mathrm{int}=515\,\mathrm{K}$ were assumed, as in \citet{Carone2023}. The model was initialized at rest with the 1D analytic pressure--temperature profile of \citet{ParmentierPF2015} above 1~bar suitable for the planet's irradiation temperature averaged over the whole planet and below 1~bar a profile from \citet{Guillot2010} consistent with the assumed $T_\mathrm{int}$. The simulation ran for 1000\,d simulation time with a dynamical timestep of $25\,\mathrm{s}$ to ensure that the temperature structure no longer evolves in the modelling domain. For cloud modelling (see Section~\ref{subsec:methods_clouds}), the snapshot after 1000\,d simulation time is used as input.

\subsubsection{RM-GCM}

The RM-GCM was originally adapted for hot Jupiter atmospheres by \citet{rauscher2010} from the University of Reading's Intermediate General Circulation Model \citep{Hoskins1975} with subsequent modifications to include radiatively active cloud feedback \citep{Roman2017,Roman2019} and ``picket-fence'' non-gray radiative transfer \citep{MalskyPF2024}. The dynamical core solves the primitive equations of meteorology on a pseudo-spectral grid, using spherical harmonics to solve the equations horizontally while the vertical equations are solved on a logarithmically-spaced sigma coordinate grid, normalized to the surface pressure.  In the models presented here, pressures range from $10^2$ bar to $10^{-5}$ bar in 65 layers. The horizontal resolution is T31, corresponding to $\sim4^{\circ}$, and the hyperdissipation is used to maintain numerical stability with a hyperdiffusion timescale of $\sim2000$\,s. Simulations were initialized with no winds and a single double-gray pressure--temperature profile \citep{Guillot2010} approximating a global average and were run for 1000 planet days.  A dynamical timestep of $\sim60$\,s was used, with radiative transfer tendencies updated every four dynamical timesteps. Picket-fence radiative transfer parameters were taken from \citet{ParmentierPF2015} with Rosseland mean opacities for $10\times$ solar composition used in place of solar composition opacities. A stellar bolometric flux of $4.2\times 10^5\,\mathrm{W\,m^{-2}}$ and an internal heat flux corresponding to $T_\mathrm{int}=500\, \mathrm{K}$ are assumed. Two versions were run, one with a clear atmosphere and another with radiatively active clouds (described further in section \ref{subsec:methods_clouds}).

\subsubsection{SPARC/MITgcm}
\label{sec:sparc}
SPARC/MITgcm \citep{ShowmanEtAl2009} uses the MITgcm dynamical core \citep{Adcroft2004} coupled to the code of 
\citet{MarleyMcKay1999}, a wavelength-dependent, plane-parallel two-stream radiative transfer code that has been widely used in the exoplanets and brown dwarf communities. Here, we use the 11 wavelength bins suggested by \citet{KatariaEtAl2013} spanning from 0.26 to 324.86~$\mu$m with eight k-coefficients for each bin.
The numerical setup of the simulation in this work is identical to the ``passive, correlated-k'' simulation in \citet{SteinrueckEtAl2023}, except that the WASP-39b planetary parameters were adopted. A horizontal resolution of C32 (equivalent to 128 $\times$ 64 grid points in the longitude and latitude directions) and 60 vertical layers spanning a pressure range from 200 bar to $2 \times 10^{-7}$~bar were used. A bottom drag and a fourth-order Shapiro filter were applied for numerical stability and convergence.   A dynamical timestep of 25\,s and a radiative timestep of 50\,s are used.

As in \citet{MayKomacekEtAl2021} and \citet{SteinrueckEtAl2023}, a hot interior boundary condition \citep{ThorngrenEtAl2019InternalTemperature} and a fixed temperature in the bottom layer were used. Based on the 10$\times$~solar metallicity grid from \citet{ThorngrenEtAl2019InternalTemperature}, the temperature at the center of the bottom-most layer at 170~bar was set to 4154~K. This is equivalent to assuming an internal heat flux equivalent to 356~K, consistent with Eq. (3) in \citet{ThorngrenEtAl2019InternalTemperature}. Molecular opacities were based on \citet{FreedmanEtAl2008, FreedmanEtAl2014} and \citet{LupuEtAl2014}, with the more recent updates described in \citet{MarleyEtAl2021Sonora}. Molecular species included as opacity sources are H$_2$O, CO$_2$, CO, CH$_4$, NH$_3$,  N$_2$, HCN, C$_2$H$_2$, C$_2$H$_4$, C$_2$H$_6$, CrH, FeH, H$_2$S,  LiCl, MgH, OCS, PH$_3$, SiO, TiO and VO. Atomic species included were Li, Na, K, Rb and Cs. Furthermore, Rayleigh scattering from H$_2$, H and He as well as collision-induced absorption from H$_2$-H$_2$, H$_2$-H, H$_2$-He, and H$_2$-CH$_4$ were included.
Bound-free and free-free opacity from H$^-$ and H$_2^+$, free-free opacity from H$_2^-$ and He$^-$, and electron scattering are further included.
The abundances for the opacity calculation were calculated at thermochemical equilibrium for a pressure--temperature grid based on the models in \citet{Gharib-NezhadEtAl2021} and the Sonora atmosphere model \citep{MarleyEtAl2021Sonora}. Rainout, in the sense that condensable species were removed from the gas mixture when their abundance exceeded the vapor pressure, was included in the chemistry calculation \citep[see ][ and references therein for detail]{MarleyEtAl2021Sonora}.

The simulation was started from a state of rest with an initial pressure--temperature profile based on the 10$\times$~solar metallicity grid from \citet{ThorngrenEtAl2019InternalTemperature}. We ran the simulation for 4,500\,d simulation time with the synthetic spectra generated from a snapshot  at the end of the simulation.

\subsubsection{THOR}
THOR is an open-source GCM developed specifically for simulations of exoplanetary atmospheres free from Solar System-centric assumptions. Various aspects of the THOR model have been successively updated in works by  \citet{mendonca_thor_2016, mendonca_2018a, mendonca_2018b}, \citet{deitrick_thor_2020, deitrick_thor_2022} and \citet{noti_examining_2023}. THOR solves the non-hydrostatic deep (NHD) Euler equations on an icosahedral grid \citep{tomita_new_2004, mendonca_thor_2016,deitrick_thor_2020,Akin2025} by design, removing many commonly used assumptions, such as the hydrostatic, traditional or shallow approximations, and admits the propagation of vertical acoustic waves. The use of an icosahedral grid in the horizontal direction \citep{tomita_shallow_2001}, with a horizontal resolution of 1$\degree$, avoids any issue of ``resolution crowding'' characterized by a convergence of grid cells along specific points on a given grid resulting in model instabilities (e.g., the singularity along the polar axis in spherical coordinates). The vertical grid is uniformly spaced in altitude with 50 levels.  To maintain numerical stability and to avoid non-physical energy build-up at the grid scale, the THOR model employs horizontal and vertical hyperdiffusion terms as well as a divergence damping operator and a sponge layer \citep{Jablonowski_2011,mendonca_2018b}. However, a basal drag is not included.   An in-depth review of how these numerical dissipation schemes affect our solutions can be found in \citet{hammond_numerical_2022}.  For the current study, the double-gray two-stream radiative transfer module \citep{deitrick_thor_2020} is used with an opacity of $\kappa_\mathrm{v} = 0.0055\, \mathrm{cm^2\,g^{-1}}$  in the visible band and $\kappa_\mathrm{th} = 0.01\,\mathrm{cm^2\,g^{-1}}$ in the thermal band. The simulation is initialized at rest with a temperature-pressure profile derived from the analytical picket-fence model at the substellar point, as in \citet{noti_examining_2023}, assuming an internal heat flux equivalent to $T_{\text{int}} = 358\,\mathrm{K}$ \citep{ThorngrenEtAl2019InternalTemperature}.  Following the recommendations of \citet{sainsburymartinez2019}, the simulation uses a hot adiabatic initial profile for the deep atmospheric layers of hot Jupiters. The simulation is run for 5000\,d simulation time, with a dynamical and radiative timestep of 100\,s. The presented results are time-averaged over the last 500\,d.

\subsubsection{The Unified Model}
\label{subsubsec:um}

The equilibrium and kinetic chemistry simulations were done using the idealised configuration of the Met Office Unified Model (UM), whose dynamical core \citep[ENDGAME,][]{Wood2014,Mayne2014a,Mayne2014,Mayne2017}, radiative \citep[SOCRATES,][]{Edwards1996,Edwards1996a,Amundsen2014,Amundsen2016,Amundsen2017} and chemistry \citep{Drummond2018,Drummond2018b,Drummond2018a,Drummond2020} components were adapted to simulate the atmospheres of hot Jupiters. ENDGAME solves the full, deep-atmosphere, non-hydrostatic Euler equations using a semi-implicit semi-Lagrangian scheme. SOCRATES solves the two-stream equations in 32 spectral bands covering 0.2–322 $\mu$m and with absorption due to \ce{H2O}, \ce{CO}, \ce{CO2}, \ce{CH4}, \ce{NH3}, \ce{HCN}, \ce{Li}, \ce{Na}, \ce{K}, \ce{Rb}, \ce{Cs}, collision-induced absorption due to \ce{H2}-\ce{H2} and \ce{H2}-\ce{He} and Rayleigh scattering due to \ce{H2} and \ce{He} \citep[see][for line list information]{Goyal2020} using the correlated-k and equivalent extinction methods. The chemistry module designed for exoplanet applications is coupled to the dynamical core and radiative transfer, and here we investigate both the chemical equilibrium and the chemical kinetics cases (see Section~\ref{subsubsec:method_kinetics} for further details regarding the chemical kinetics). The UM uses a constant angular grid, with the simulation having 144 longitudinal and 90 latitudinal grid points resulting in a horizontal resolution of \ang{2.5} in longitude by \ang{2} in latitude and 66 vertical levels equally spaced in height covering pressures from $\sim$ 200 bar to $\sim$ $10^{-5}$ bar. The atmosphere is assumed to be aerosol-free with 10$\times$solar metallicity, and was run for 1000\,d simulation time until the upper atmosphere ($\leq$\SI{e5}{\pascal}) had reached a pseudo-steady state dynamically, radiatively and chemically. The model employs a sponge layer covering the upper 1/4 of the computational domain with a damping coefficient of 0.15, as well as a second-order longitudinal filter with a diffusion coefficient of $K=0.158$. A discussion of the diffusion and damping can be found in \citet{Mayne2014} and \citet{christie2024}. The UM simulations were initialised at rest with the dayside-average pressure--temperature profile from the 1D radiative-convective-chemistry model ATMO \citep[e.g., ][]{Tremblin2015,Drummond2016} which was run assuming chemical equilibrium for the chemical species present in the \citet{Venot2012} chemical network. The internal heat flux was assumed to be equivalent to $T_\mathrm{int}=100\,\mathrm{K}$ in both ATMO and the UM simulations, with no frictional drag applied at the bottom boundary in the UM.   The UM simulations  were run using dynamical and radiative timesteps of $30\,\mathrm{s}$ and $150\,\mathrm{s}$, respectively.

\subsection{Disequilibrium Chemistry} \label{subsec:methods_2Dchem}
To better ascertain what impact disequilibrium chemistry may play in the observed limb asymmetries, we adopt two approaches.  The first is to simulate chemical kinetics directly in the UM GCM, capturing the interaction of transport and thermochemistry.  The second is to post-process each of the GCM simulations using the 2D VULCAN photochemical model, allowing the photochemistry to be followed.  Each approach is expanded upon below.

\subsubsection{Transport-induced Disequilibrium Thermochemistry in 3D with the Unified Model}
\label{subsubsec:method_kinetics}
\label{subsubsec:um_diseq}
The UM gas phase abundances can be calculated under the assumption of chemical equilibrium, through minimization of the Gibbs energy, or through direct calculation of the time-dependent reaction rates in a given chemical network, termed chemical kinetics \citep[described in][]{Drummond2020}. In this work we track the 30 species found in the \citet{Venot2019} reduced C/N/O/H chemical network, which are advected and interact with the radiative transfer, solving for the chemical state every 3750\,s (125 dynamical timesteps). For the chemical kinetics, the abundances are initialized to their equilibrium values at the beginning of the simulation and then integrated using the 181 reversible reactions from \citet{Venot2019}. As the model does not include sub-grid eddy or molecular diffusion, or photochemical reactions, the only mechanism the only mechanism that can drive abundances away from their local equilibrium values is atmospheric transport \citep[e.g.,][]{ZamyatinaEtAl2024WASP-96b}.  Except for the approach to the solution for the chemical abundances, the configuration of the chemical kinetics simulation is identical to that of the equilibrium chemistry simulation. The two simulations nevertheless developed slightly different temperature structures, through the radiative feedback of their differing chemical compositions. These temperature differences are small (about 6 K on average, and 35 K as maximum, at pressures below 0.1 bar), so the differences in absorber abundances between the two simulations, and hence in their transmission spectra (Section 4.1), arise predominantly from transport-induced departures from local equilibrium.

\subsubsection{Photochemistry with the 2D Kinetics Model VULCAN}
Photochemistry affects the abundances of multiple major absorbers in the atmospheres of hot Jupiters. Most notably, it leads to the production of detectable amounts of \ce{SO2} \citep{Tsai2023a} and the photochemical destruction of \ce{CH4} at low pressures \citep[e.g.,][]{MosesEtAl2011}. Yet, photochemistry is not included in the UM chemical kinetics simulations.

To explore how circulation interacts with thermochemical and photochemical processes, we employ the 2D photochemical model VULCAN to simulate the gas-phase composition in the equatorial region. VULCAN includes both vertical mixing and horizontal transport on a longitude-pressure plane using velocity fields from the contributing GCMs as input, and it has been benchmarked and described in detail in \cite{Tsai2024}. Here, we briefly summarize the model setup. The 2D photochemical model solves the continuity equation
\begin{linenomath*}
\begin{equation}
\frac{\partial n_i}{\partial t} = {\cal P}_i - {\cal L}_i - \frac{\partial \phi_{i,z}}{\partial z} - \frac{\partial \phi_{i,x}}{\partial x},
\label{eq:master}
\end{equation}
\end{linenomath*}
where $n_i$ is the number density of species $i$, ${\cal P}_i$ and ${\cal L}_i$ are the chemical production and loss rates of species $i$, and $\phi_{i,z}$, $\phi_{i,x}$ are the vertical and horizontal transport fluxes, respectively. The vertical flux includes transport by molecular diffusion and eddy diffusion, expressed as 
\begin{equation}
\phi_{i,z} = - K_{\textrm{zz}} N \frac{\partial X_{i}}{\partial z} - D_{i} \left[ \frac{\partial n_{i}}{\partial z} + n_{i}(\frac{1}{H_{i}} + \frac{1+\alpha_T}{T}\frac{d T}{d z}) \right],
\label{eq:phiz}
\end{equation}
where $K_\mathrm{zz}$ is the mixing rate inferred from the GCM, $D_{i,z}$ is the molecular diffusion coefficient, $N$ is the total number density, $X_{i}$ is the volume mixing ratio of species $i$, $H_i$ is the scale height of species $i$, and $\alpha_T$ is the thermal diffusion factor. For numerical stability, vertical advection is not included here. The horizontal transport by the zonal wind $v_x$ is 
\begin{equation}
\phi_{i,x} = - n_{i,x} v_x,
\label{eq:phix}
\end{equation}
evaluated on isobaric surfaces.

2D VULCAN adopts the temperature and wind profiles obtained from the GCM output introduced in Section \ref{subsec:methods_gcms}. These profiles are averaged over an equatorial latitudinal band of $\pm$ 30$^{\circ}$, where the circulation is characterized by an equatorial jet, allowing the system to be represented by a 2D framework. We use the same chemical network as \cite{Tsai2023a}, except for a minor update on the rate constants associated with \ce{H2S}\footnote{\url{https://github.com/exoclime/VULCAN/blob/master/thermo/SNCHO_photo_network_2024.txt}}. This 2D framework simplifies the 3D global circulation and excludes chemical feedback on the thermal structure; however, the 2D photochemical model offers insights into global chemical transport not readily accessible through 1D photochemical or 3D GCMs. 

A challenge when modeling photochemistry with GCMs is that the upper boundaries used in GCMs typically do not extend to sufficiently low pressures for photochemical models. 2D VULCAN uses a top boundary of 10$^{-8}$~bar, while the top boundary of GCMs in this work ranges from $10^{-4}$ to $2\times10^{-7}$~bar. For regions above the top boundary of the GCM, we assume that temperature and winds remain constant with height, with the values set to those at the top boundary of the GCM. In Appendix \ref{sec:photochem_upper_boundary}, we demonstrate that this assumption does not qualitatively change our results.

\subsection{Condensate Clouds}

We similarly adopt two approaches to cloud modelling to investigate what impact they may have on the observed limb asymmetry.  Our first approach is to post-process the ExoRad GCM output using the IWF~Graz 1D microphysical model.  The second is to model simplified, radiatively-coupled clouds in the RM-GCM to observe what impact they have on the thermal structure.  Each approach is outlined below.

\label{subsec:methods_clouds}
\subsubsection{Post-processed Clouds with the IWF Graz Microphysics Model}
\label{ssubsec:IWF_clouds}
\label{subsec:IWF_cloud_description}
Due to the computational overhead of coupling cloud microphysics to a GCM \citep[see, e.g.,][]{LeeEtAl2016,lines2018},  we instead post-process the outputs of ExoRad with the IWF~Graz microphysical kinetic cloud formation model, which, when used in tandem, we denote as the ExoRad/IWF Graz model.  The cloud model is coupled to equilibrium gas-phase chemistry, which includes the processes of nucleation, bulk growth, evaporation, and gravitational settling of cloud particles, as well as element consumption and replenishment \citep{Woitke2003_Paper2,Woitke2004_Paper3,Helling2006_Paper5,Helling2008_Paper1_new}. We consider cloud condensation nuclei to form from four species (\ce{TiO2}, SiO, NaCl, KCl) and compute the nucleation rates for each species using modified classical nucleation theory while condensation of 16 condensate species (\ce{TiO2}[s], \ce{Mg2SiO4}[s], \ce{MgSiO3}[s], MgO[s], SiO[s], \ce{SiO2}[s], Fe[s], FeO[s], FeS[s], \ce{Fe2O3}[s], \ce{Fe2SiO4}[s], \ce{Al2O3}[s], \ce{CaTiO3}[s], \ce{CaSiO3}[s], \ce{NaCl}[s], \ce{KCl}[s]) is considered through 132 gas-surface reactions. A replenishment 
timescale, $\tau_{\rm mix}$, is used to simulate
replenishment of gas species through upward mixing 
from the deep, cloud-free atmosphere into the cloud forming regions.  
Together with 11 element conservation equations, a set of 19 moment equations are solved \citep[see, e.g.,][]{Helling2006_Paper5} from which cloud properties (mean cloud particle size, cloud particle number density, mixed material compositions of cloud particles) are derived.

As input for the IWF~Graz model, 1D pressure, temperature, and vertical velocity profiles were extracted from the cloud-free ExoRad simulation (Sec. \ref{subsubsec:methods_gcm_ExpeRT/MITgcm}) to provide the atmosphere structure in a similar hierarchical approach to what is used in \cite{Helling2021_WASP_Comparison,Helling2023_Grid}. The 1D profiles were extracted at longitudes $\pm 75^{\circ}$, $\pm 90^{\circ}$, and $\pm 105^{\circ}$ and at latitudes $0^{\circ}$, $23^{\circ}$, $45^{\circ}$, $68^{\circ}$, and  $86^{\circ}$ to generate the spectra. The replenishment timescale $\tau_{\rm mix}$ is derived from the standard deviations of the ExoRad GCM vertical velocities as in Appendix B in \cite{Helling2023_Grid}. \cite{Samra2023_WASP96b} find that increasing $\tau_{\rm mix}$ by a factor of $100\times$ compared to the GCM-derived values leads to better agreement with observations for WASP-96b, as using the GCM-derived value leads to an overabundance of clouds. The same approach is used here for WASP-39b.  
The cloud model assumes an undepleted gas abundance of 10$\times$~solar.

\subsubsection{Parametrized Radiatively Active Clouds in the RM-GCM}
\label{subsec:RMGCM_cloud_description}
In the RM-GCM, clouds are treated according to the model of \citet{Roman2019}, with the updates detailed in \citet{Roman2021} and \citet{MalskyPF2024}. In contrast to the post-processing approach of the Graz model described in 2.3.1, the RM-GCM determines the presence or absence of each cloud species as the simulation progresses based on local temperatures and pressures with the clouds condensing and dissipating instantaneously as pressure and temperature conditions evolve. When present, clouds absorb and scatter light, allowing them to actively alter the radiative transport and resulting thermal structure. The default cloud prescription includes eight cloud species; KCl[s], Cr[s], SiO$_2$[s], Mg$_2$SiO$_4$[s], VO[s], Ca$_2$SiO$_4$[s], CaTiO$_3$[s], and Al$_2$O$_3$[s]. Clouds are present anywhere in the atmosphere where pressure--temperature conditions allow, based on equilibrium-chemistry condensation curves for a given composition from \citet{Mbarek2016}. 
To approximate a balance between vertical mixing and gravitational settling, a maximum thickness (in units of vertical layers) is prescribed. This is applied on a per-species and per-column basis: for each column, each species of cloud is only allowed to be present up to some number of layers above the lowest intersection of the condensation curve and the temperature profile. Past work with cloudy models using the RM-GCM has interrogated limb asymmetry in the extremes of very thick and thin clouds \citep[][]{Roman2021, triantafillides2025out}. This work showed that in the temperature regime of WASP-39b, very thick prescriptions eliminate all gas-phase features, while very thin ones contribute negligibly to the transmission spectrum. We chose an intermediate thickness of 35 layers (about 4 orders of magnitude in pressure) to allow silicate clouds to impact the temperature structure at low pressures without completely obscuring the gas-phase features.
In the top four permitted layers, the cloud mass is gradually tapered off to 0 (multiplied by 0.01, 0.03, 0.1, and 0.3). This tapering was necessary for numerical stability due to the supersolar metallicity (and therefore increased cloud-forming mass) of WASP-39b.
The cloud mass that condenses is set by the abundance of the stoichiometrically limiting constituent of each species, assuming solar abundance ratios scaled to a chosen metallicity. A constant 10\% of that available mass condenses where conditions allow a cloud to be present.  This reduction is necessary to improve numerical stability in the model \citep[and is common in GCMs with silicate cloud radiative feedback, e.g.][]{Mehta2025} and approximates inefficient condensation.  \citet{Roman2021} showed that this practical adjustment, in general, reduced the magnitude of cloud effects, but still yielded similar qualitative effects in capturing the impact of clouds. Overlap in the composition of cloud species is neglected, since none of the included clouds share a limiting constituent and only 10\% of the limiting mass is used. Particle sizes are prescribed as a function of pressure, set to 0.1 $\mu$m at P $\leq$ 10 mbar, and increasing linearly with pressure to 80 $\mu$m at the model bottom boundary of 10$^2$ bar following \citet{Roman2021}. Conditions for cloud formation and the cloud radiative properties are evaluated at each radiative timestep. Cloud absorption and scattering properties in each radiation band are interpolated from a pressure--temperature grid, pre-calculated using Mie theory and assuming spherical grains of a single radius at each pressure level, as described in \citet{MalskyPF2024}. When multiple species of cloud are present at the same point, the scattering parameters (e.g. the single scattering albedo and asymmetry parameter) are taken to be the optical-depth weighted average of the scattering parameters of each component cloud \citep[see Equations 11 \& 12 in][]{Malsky2021}.

It is worth noting that the above assumptions, combined with the global temperature regime of WASP-39b, restrict the most radiatively important clouds (SiO$_2$[s] and Mg$_2$SiO$_4$[s]) to be horizontally homogeneous in particle size and mass mixing ratio. These simplifications allow us to efficiently compute changes in the 3-D temperature structure induced by these prescribed clouds, including limb temperature differences.

\subsection{Photochemical Hazes}
\label{subsec:hazes_description}
We modeled photochemical hazes within SPARC/MITgcm as passive (i.e., not radiatively coupled) tracers with a constant particle size following \citet{SteinrueckEtAl2021}. Hazes are produced at low pressures, with a vertical haze production profile that is a log-normal distribution in pressure, with a median of 2~$\mu$bar and a standard deviation of 0.576. Horizontally, the column-integrated haze production rate peaks at the substellar point and scales with the cosine of the angle of incidence of the starlight. Deeper in the atmosphere, at pressures $> 0.1$ bar, hazes are destroyed by a sink term. We considered particle sizes varying from 1~nm to 1000~nm, but only results for the 30~nm case---representative of the small particle regime---are shown here as larger particle sizes are unlikely based on microphysical modeling \citep{LavvasKoskinen,ArfauxLavvas2022}, and qualitatively provided a worse match to the NIRSpec PRISM data. With a passive tracer model, the haze production rate can be scaled by an arbitrary factor after running the simulation. We chose a column-integrated haze production rate of $2.5\times 10^{-12}$ kg~m$^{-2}$~s$^{-1}$ at the substellar point as our nominal case, which resulted in the best match to the observed limb-averaged transmission spectrum, but also include larger haze production rates for the purpose of illustration.

\subsection{Generating Spectra}
\label{subsec:spectra_description}

To compare the results from the GCM simulations to the observations, we generate synthetic transmission spectra for each of the GCM outputs, following the procedure of \citetalias{espinoza2024inhomogeneous}. We perform extinction-only, ray-striking radiative transfer with the 3D post-processing code outlined in \cite{kempton2012constraining}, with the morning and evening spectra generated by simply mirroring the morning (evening) limb onto the evening (morning) limb (i.e., constructing ``morning-only'' and ``evening-only'' atmospheres).  As the rays traverse physical lines of sight, GCMs that employ pressure grids in the vertical direction (i.e., all but THOR and the UM) have their vertical structure mapped onto equally spaced altitude grids with the inner boundary taken to be at 1 bar.  For THOR and the UM, which use an altitude-based vertical grid, we use the highest altitude where the pressure always exceeds 1 bar as the inner boundary. 

The opacity sources included in generating the spectra are $\rm H_2O$, $\rm CO$, $\rm CH_4$, $\rm CO_2$ , $\rm C_2H_2$, $\rm SO_2$, and $\rm NH_3$. The opacity inputs are generated using opacity sampling with the {HELIOS} {k-table} module \citep{GrimmHeng2015HELIOS-K,malik2017helios,grimm2021helios} at a resolution of 3000 from 0.5 to 21 microns, and the line lists used can be found in Table \ref{tab:opacitydata} in Appendix \ref{sec:opacitydata}. Then, we degrade our synthetic spectra to a resolution of 100 following the same procedure as \cite{line2021solar}.

For the UM, we use GCM-output gas abundances to determine the local volume mixing ratios (VMRs) of each gas species considered. For all other GCMs, we determine the local VMRs by interpolating FastChem \citep{stock2018fastchem, stock2022fastchem} equilibrium chemistry tables as a function of local temperature and pressure. These tables are calculated at $10\times$ solar metallicity to agree with the metallicity assumed in the GCMs.

Unlike the thermochemical models,  where the three-dimensional chemical structure of the atmosphere is captured directly in the GCM, the photochemical calculations are fundamentally two-dimensional, as they are only intended to model transport near the equator. Thus the spectral post-processing requires assumptions about how to extend the chemical abundances in the meridional dimension. We consider two limiting cases: in the first, we assume that the abundances are constant in latitude and thus photochemistry is modelled throughout the atmosphere (we refer to this version as ``\BOne'').  In the second, we assume that the photochemical model only represents the state of the atmosphere within 30 degrees of the equator, approximating the behavior of an equatorial jet, with the remainder of the atmosphere assumed to be in chemical equilibrium (we refer to this version as ``\BTwo'').

The spectral post-processing of the RM-GCM clouds follows \cite{triantafillides2025out}. The optical depth in each model layer $\Delta\tau_{\mathrm{cloud,\lambda, i}}$ is computed consistently with the GCM \citep[see][equation 3]{MalskyPF2024}, with condensate extinction efficiencies calculated from a Fortran conversion\footnote{\url{https://github.com/ELeeAstro/Rosseland_Clouds}} of \texttt{LX-MIE} \citep{Kitzmann2018}. The optical depth of each model layer is then converted to an attenuation coefficient for each cloud species, $\alpha_{\rm cloud, i}$, using 

\begin{equation}
\label{eq:ext}
    \alpha_{\rm cloud,\lambda, i} \, [m^{-1}]= \frac{\Delta\tau_{\mathrm{cloud,\lambda, i}}}{\Delta s}
\end{equation}
\noindent for a layer thickness $\Delta s$. We sum the $\alpha_{\rm cloud, \lambda, i}$ to arrive at the total cloud attenuation coefficient across all 8 species (KCl, $\rm Cr_2O_3$, $\rm SiO_2$, $\rm Mg_2SiO_4$, VO, $\rm Ca_2SiO_4$, $\rm CaTiO_2$, and $\rm Al_2O_3$):

\begin{equation}
\label{eq:cloud}
    \alpha_{\rm cloud,\lambda,  total} = \sum_{i=1}^{8} \alpha_{\rm cloud, \lambda, i}. 
\end{equation}

Our approach does not account for the depletion of gas-phase species by cloud formation. Rather, gas-phase abundances are set entirely by equilibrium chemistry (except for the photochemical cases previously described).

We treat clouds differently when post-processing the ExoRad/IWF Graz outputs. The post-processing of the ExoRad simulations with the IWF~Graz cloud model outputs the cloud scattering and absorption coefficients calculated on 5 unique latitudes (+ 4 mirrored from the northern to the southern hemisphere), 6 longitudes, and interpolated onto 50 pressure points. The cloud species included are \ce{TiO2}[s], \ce{Mg2SiO4}[s], \ce{MgSiO3}[s], MgO[s], SiO[s], \ce{SiO2}[s], Fe[s], FeO[s], FeS[s], \ce{Fe2O3}[s], \ce{Fe2SiO4}[s], \ce{Al2O3}[s], \ce{CaTiO3}[s], \ce{CaSiO3}[s], \ce{NaCl}[s], \ce{KCl}[s] and unlike the RM-GCM, mixed-composition grains are considered. We calculate cloud scattering and absorption coefficients using Mie scattering and Landau-Lifshitz-Looyenga effective medium theory \citep{LandauLifshitz1960,Looyenga1965}
The cloud scattering and absorption coefficients are combined to yield mass attenuation coefficients (in m$^2$/kg), which are then interpolated to the gas opacity wavelength grid. Finally, they are linearly interpolated from the coarser grid of the cloud model to the finer post-processing grid as a function of latitude, longitude, and pressure, and they are scaled by the local gas density to yield units of inverse length. Again, depletion of gas-phase species by cloud formation is neglected when generating spectra.

We similarly include the optical effects of hazes when post-processing the SPARC/MITgcm simulations. Haze particle extinctions are calculated with Mie theory \citep[using PyMieScatt;][]{sumlin2018retrieving} assuming a particle size of 30~nm  and a refractive index of soot \citep{LavvasKoskinen}.

For both the RM-GCM and ExoRad/IWF~Graz cloud models as well as the SPARC/MITgcm haze model, the aerosol opacity is then added to the gas opacity to compute the total opacity in a given cell.

\section{Temperature structure of clear-atmosphere models} \label{sec:temperaturestructure}

\begin{figure*}
    \centering
    \includegraphics{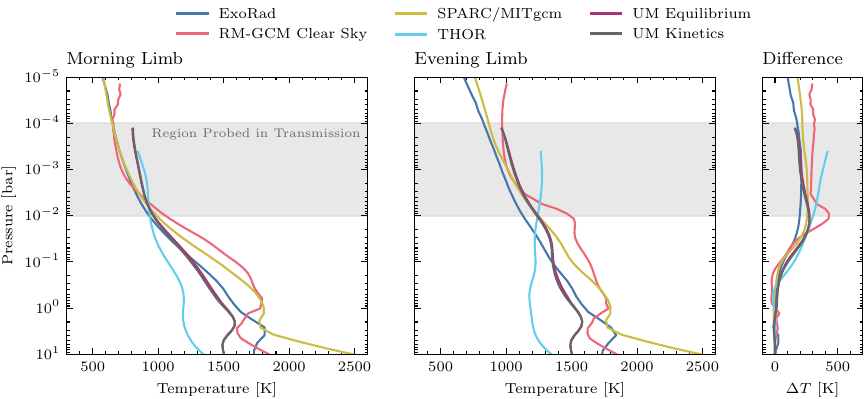}
    \caption{Pressure--temperature profiles at the morning (left panel) and evening (center panel) terminators for each clear-sky GCM. The temperature difference between the limbs are shown in the right panel, with a positive difference indicating a hotter evening terminator.  The profiles are averaged between the latitudes of $-30\degree$ and $+30\degree$. While the absolute temperatures at the terminator vary substantially between simulations done using different GCMs due to a variety of model differences, the difference between morning and evening terminator remains remarkably consistent between all clear-atmosphere models.}
    \label{fig:limb_pt}
\end{figure*}

\begin{figure*}
    \centering
    \includegraphics{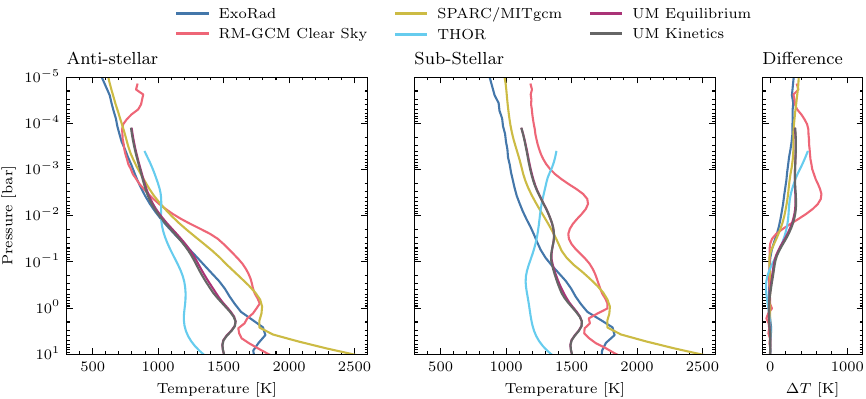}
    \caption{Pressure--temperature profiles at the anti-stellar (left panel) and substellar (center panel) points for eachclear-sky GCM. The temperature difference is shown in the right panel, with a positive difference indicating a hotter dayside. The profiles are averaged between the latitudes of $-30\degree$ and $+30\degree$. The clear-sky RM-GCM (pink-red) has the hottest dayside, likely due to the use of picket-fence radiative transfer. }
    \label{fig:daynight_pt}
\end{figure*}

\begin{figure}
    \centering
    \includegraphics{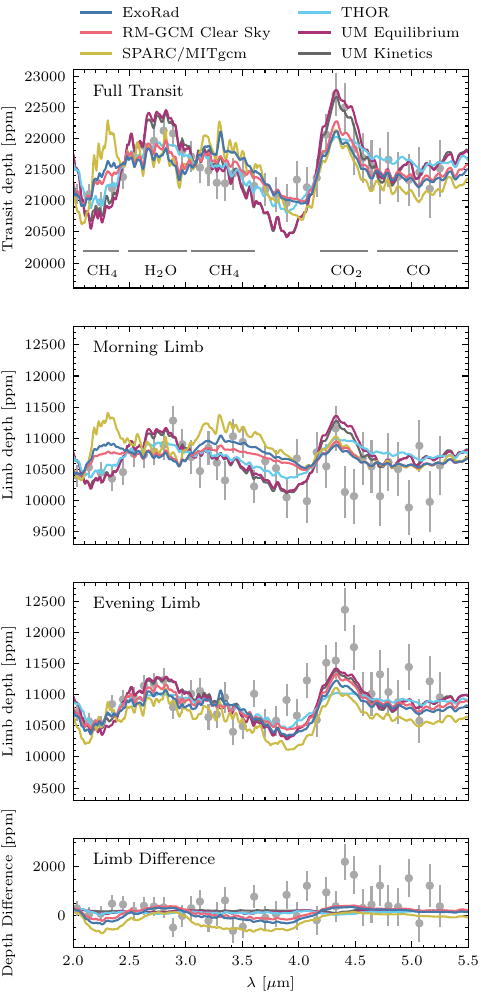}
    \caption{Transmission spectra for the clear sky simulations with the observational data in grey.  The full transmission spectra are in the top panel with the individual limbs shown in the middle panels.   The difference between the limbs, with a larger evening limb being positive, is shown in the bottom panel.  For several of the models the largest morning--evening difference lies at 3.5 microns due to the strong \ce{CH4} features on the morning terminator, which is not seen in the observations. In contrast, the observations show a large morning--evening difference in the size of the \ce{CO2} feature, which is not borne out by the models.}
    \label{fig:trans_clear}
\end{figure}

We begin by considering the simulations without clouds,  hazes, or disequilibrium chemistry to better understand the baseline which the cloud and haze models build upon and to what extent these preliminary clear sky atmospheres can reproduce observations.

\subsection{Deep Atmosphere Structure}

It is important to first note that as these simulations use a number of different internal heat fluxes, initial conditions and levels of numerical dissipation, as well as varied integration times, the behavior in the deep atmosphere is both expected and observed to vary greatly between models due to the long integration times required to approach convergence (Fig. \ref{fig:limb_pt}, left and center panel; see also \citealt{sainsburymartinez2019}). However, as the deep atmosphere is horizontally uniform and shows no evidence of asymmetries and contributes minimally to the transmission spectra, we do not focus on it except to point out that the differences between models can be $\sim 1000\,\mathrm{K}$ and any inferences relying on results from this region of the atmosphere should be viewed with skepticism.  We further note that while we do not discuss the global temperature structure or velocities of the GCMs in depth in this section, zonal wind speeds and temperature structure on isobars are shown in Appendix \ref{sec:circulation} for reference.

\subsection{Terminator Temperature Differences}
Higher in the atmosphere, temperature asymmetries are seen along the terminators above $\sim 0.1\,\mathrm{bar}$ with a peak in the temperature asymmetry between the two limbs around $0.01$ bar (Fig. \ref{fig:limb_pt}, right panel). The magnitude of the peak limb temperature differences varies between simulations, with the clear RM-GCM simulation exhibiting a 400 K difference between the morning and evening terminators, while the UM and SPARC/MITgcm simulations peak at a somewhat lower temperature $\sim 300\,\mathrm{K}$ and ExoRad peaks at $200\,\mathrm{K}$.   

The larger limb asymmetry seen with the RM-GCM is driven, in part, by the hotter dayside temperatures, reaching $1700\,\mathrm{K}$ at P $\sim 4\times 10^{-3}\,\mathrm{bar}$, whereas the other GCMs have substellar temperatures of 1000~K to 1300~K at similar pressures (see Fig. \ref{fig:daynight_pt}). This difference may simply be due to the use of picket fence radiative transfer in RM-GCM with the other GCMs using correlated-k multiband or dual-grey radiative transfer. It may also be due to the specific picket-fence setup used in the RM-GCM simulations wherein the Rosseland mean opacities are computed for a $10\times$ solar metallicity atmosphere but the picket fence parameters for the visible bands, $\gamma_{\mathrm{v},i}$, are assumed to be fixed with their values set by the $1\times $ solar fits found in \citet{ParmentierPF2015}.  

The THOR simulation has the coolest deep atmosphere of any of the simulations and exhibits an increasing temperature asymmetry between the two limbs with decreasing pressure.  THOR's temperature profiles at each limb differ from other GCMs, with the evening limb being roughly isothermal with $T \sim 1200\,\mathrm{K}-1300\,\mathrm{K}$.  This different behavior is likely due to the use of double-gray radiative transfer unlike the remaining GCMs and the greater boundary pressure at the top of the model.  The temperature differences, however, remain relatively consistent with the other GCMs. 

Across all models, the limb temperature differences at $P < 1$\,mbar generally range from 100 to 300\,K. RM-GCM is the outlier, reaching $\sim$400\,K near 0.01\,bar.

\subsection{Impact on Transmission Spectra}
Each model was post-processed as outlined in Section \ref{subsec:spectra_description}, and the results are shown in Figure \ref{fig:trans_clear}.  Due to the arbitrary normalization of the atmospheric altitudes, relative to the planet radius, in the post-processing a linear shift in transit depth is applied to each model. The adjustments range from 600 to 1000~ppm and are calculated through least-squares minimization between each model and the full-transit data.  These shifts, computed from the full transit, are applied to the individual limb spectra to maintain the limb asymmetries in the individual GCMs.  The full transit spectrum (Fig. \ref{fig:trans_clear}, top panel) exhibits a prominent \ce{CO2} feature at 4.3~$\mu$m. The exact contrast between the peak of this feature and the opacity window at 3.9 $\mu$m is difficult to reproduce as it depends strongly on the relative abundances of \ce{CO2} and \ce{CH4} within the individual clear-sky simulations.  The UM simulations, for example, lack methane relative to other simulations and thus exhibit the largest contrast while the SPARC/MITgcm, ExoRad, and RM-GCM clear-sky simulations have increased methane contributions and thus smaller contrasts.  As all these models lack clouds and photochemistry and assume the same elemental abundances, the differences in the relative abundance of \ce{CH4} and \ce{CO2} are likely due to the temperature differences resulting from model choices.  For example, the temperatures on the morning limb of the UM and THOR simulations in the region probed by transmission spectroscopy -- the shaded gray region in Figure \ref{fig:limb_pt}, left panel -- are warmer than other simulations, consistent with the reduction in methane absorption seen in their spectra.  Tuning of parameters within individual GCMs may alter temperatures on the limbs and improve the agreement; however, the introduction of other physical processes, such as photochemistry (Section \ref{subsec:vulcanresults}) and clouds (Section \ref{sec:clouds}), may also impact the abundances and provide additional sources of opacity. 

When splitting the features between limbs, the \ce{CO2} feature appears primarily in the observations of the evening limb (compare Fig. \ref{fig:trans_clear}, middle panels). In the observations, the \ce{CO2} feature on the evening terminator is $\sim2000$~ppm larger than on the morning terminator. This asymmetry in the feature is not reproduced by any of the cloud-free, equilibrium-chemistry GCM synthetic spectra.  In the simulations, instead, \ce{CO2} abundances remain relatively homogeneous across the limbs with the transit asymmetry that is found  at $\sim 4.3\,\mathrm{\mu m}$ being primarily due to temperature differences and not the \ce{CO2} abundance.  The ExoRad and RM-GCM simulations do have transit asymmetries at 4.3~$\mu$m favoring a larger effective evening radius; however, these asymmetries are smaller than the observed asymmetry with the evening terminator only $\sim400$~ppm larger than the morning terminator. The amplitude of the full transit \ce{CO2} feature as well as the asymmetry may be influenced by the C/O ratio and the metallicity, both directly through the changes in the elemental abundance resulting in more \ce{CO2} at higher metallicities, and indirectly through the changing thermal structure along the limbs. However, the picture may also be complicated by clouds, which are more thoroughly explored in Section \ref{sec:clouds}. None of the models presented here tested the impact of varying metallicity and/or C/O ratio, making this an interesting avenue for future investigation.

Most of the synthetic spectra show excess absorption around 3.5 microns favoring a larger morning limb, opposite of what is found in the observations. This is due to the higher equilibrium methane abundances resulting from the cooler temperatures on the morning limb. This effect is especially prominent for the SPARC/MITgcm spectra (see the temperatures in Fig. \ref{fig:temperatures_gcms} in Appendix \ref{sec:circulation}).  A hotter morning limb would reduce the methane abundance responsible for the asymmetry while maintaining the assumption of chemical equilibrium. Were instead chemical kinetics or photochemistry included within the models, the methane abundance could be reduced through transport-induced quenching as large-scale motions carry gas with a reduced methane abundance from the deeper atmosphere and the hotter day side or through photochemical dissociation of methane at the top of the atmosphere.  These possibilities are explored further in Section \ref{sec:disequilib}.

\section{Disequilibrium Chemistry} \label{sec:chemistry}
\label{sec:disequilib}

The clear-sky models in the previous section assumed the atmosphere was in thermochemical equilibrium.  We now investigate how departures from equilibrium may influence the limb asymmetry, first examining coupled chemical kinetics in Section \ref{sec:um} and then photochemical post-processing in Section \ref{subsec:vulcanresults}.

\subsection{3D Chemical Kinetics (Unified Model)}
\label{sec:um}

To understand any potential limb asymmetry resulting from departures from chemical equilibrium, simulations were done with the UM using both equilibrium chemistry and chemical kinetics setups, as outlined in Sections \ref{subsubsec:um} and \ref{subsubsec:um_diseq}. Previous work using the UM demonstrated that the relaxation of the assumption of chemical equilibrium made little difference to the synthetic transmission spectrum for the hottest and coolest planets in their study, with the greatest impact seen for HD~189733b with an equilibrium temperature near 1200~K \citep[$\sim$150\,ppm; ][]{Zamyatina_2023}.  WASP-39b has a similar equilibrium temperature to HD~189733b, where the balance between the chemical, radiative and advective timescales results in departures from equilibrium which are potentially significant enough to be observed, although differing metallicity and other parameters between the models may yield divergent results.
\subsubsection{Chemical Abundances}
\begin{figure*}
    \centering
    \includegraphics{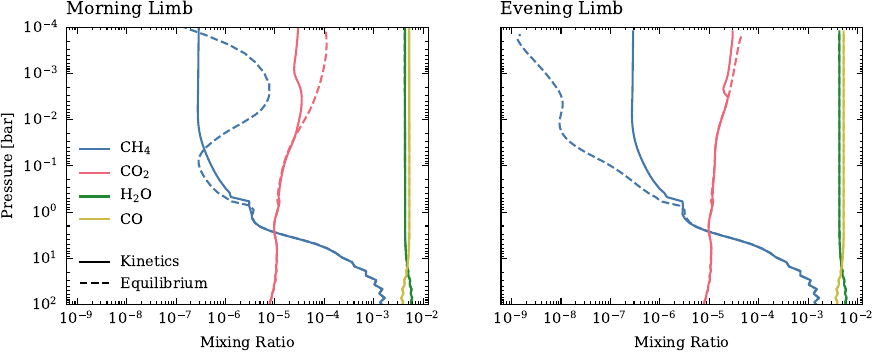}
    \caption{Equatorial chemical profiles for the UM equilibrium (dashed lines) and kinetics (solid lines) simulations at the morning (left) and evening (right) terminators. While \ce{CH4} strongly deviates from chemical equilibrium in the kinetics simulation, \ce{CO2} stays relatively close to its equilibrium abundance.  }
    \label{fig:limb_chem}
\end{figure*}

\begin{figure*}
    \centering
    \includegraphics{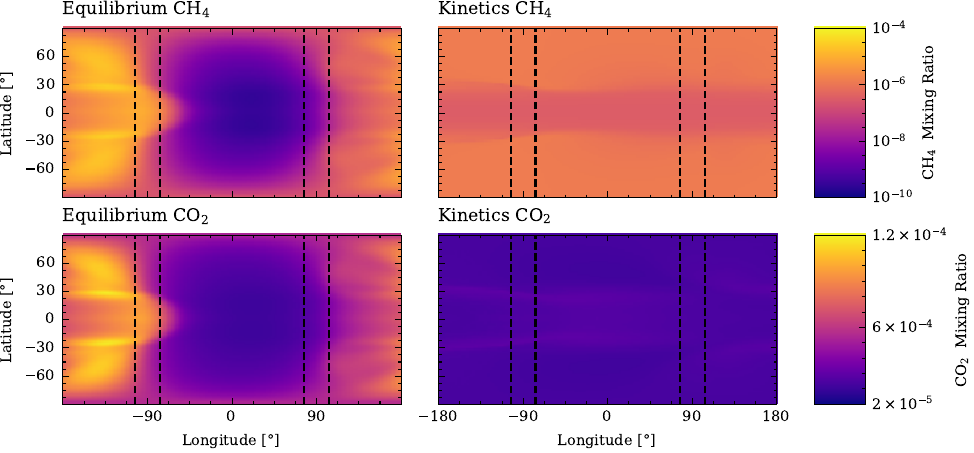}
    \caption{\ce{CH4} (top row) and \ce{CO2} (bottom row) volume mixing ratios at $10^{-3}$ bar for the UM equilibrium (left column) and chemical kinetics (right column) simulations. The plots are oriented such that the substellar point is in the center of the panel, and the morning and evening terminators are at longitudes of $-90\degree$ and $+90\degree$, respectively. The black dashed lines indicate the opening angle along the terminator, computed following \citet{Wardenier2022}. Atmospheric transport efficiently homogenizes the abundances of both species. Note that the \ce{CH4} mixing ratio is presented on a logarithmic color scale while the \ce{CO2} mixing ratio is on a linear color scale.}
    \label{fig:chem_mbar}
\end{figure*}

The simulations demonstrate that the abundances of the important absorbers \ce{CH4} and \ce{CO2} indeed exhibit substantial deviations from chemical equilibrium. Figure \ref{fig:limb_chem} shows the vertical abundance profiles on the two limbs of the simulations, for both the equilibrium and kinetics simulations. The abundances of \ce{CO_2} and \ce{CH_4} vary by an order of magnitude or two, respectively, between the equilibrium and kinetics case. For the kinetics case, the abundances are similar at both limbs, but vary by up to three orders of magnitude for \ce{CH_4} in the equilibrium case. Figure \ref{fig:chem_mbar} shows the abundances of \ce{CO_2} and \ce{CH_4} at a pressure of $10^{-3}$ bar for the equilibrium and kinetics simulations. Again, there are stronger horizontal abundance variations in the equilibrium chemistry case.  While the kinetics simulations result in both a reduction in and a homogenization of the methane abundance at low pressures, the detailed quenched methane abundance is known to depend on the temperatures in the deep atmosphere \citep{fortney2020}.  The deep atmosphere temperatures can be impacted by radiatively-coupled clouds, which are not included in the UM chemical kinetics simulations, as well as the assumed intrinsic temperature $T_\mathrm{int}$, which is taken to be $100\,\mathrm{K}$ in the chemical kinetics simulations. While we do not explore these aspects within the UM itself, we note that the variation in the quenched CH$_4$ abundances between different GCMs in the 2D VULCAN simulations in Section \ref{subsec:vulcanresults}  provides a good sense of the magnitude of this effect.

\subsubsection{Impact on Transmission Spectra}
Comparing the resulting spectra to the observed full-transit spectrum, the observations of WASP-39b are consistent with both the chemical equilibrium and chemical kinetics simulations of the UM to the same degree (Fig. \ref{fig:trans_clear}, top panel). It is important to note here that the simulations are identical in setup, aside from the approach to the gas phase chemistry. The difference between the models ($\sim$100\,ppm) is below the uncertainty of the observations.  Thus, our observations are not able to differentiate between the two synthetic transmission spectra for what is potentially one of the optimal cases. However, we note that this result may be sensitive to modeling choices within the GCM and the particular GCM used. The UM simulations are the hottest at pressures below $\lesssim 10^{-3}$ bar relative to other GCMs in this study and thus have the lowest methane abundances. In comparison, the simulations with the coldest temperatures at low pressures (SPARC/MITgcm, ExoRad) exhibit strong methane features in equilibrium chemistry and may thus exhibit larger spectral changes if disequilibrium thermochemistry is introduced. However, these models currently are not capable of simulating chemical kinetics.

Any difference between the limbs in the synthetic transmission spectra is small (Fig. \ref{fig:trans_clear}, bottom panel). The asymmetry in the observations resulting from the strong evening \ce{CO2} feature at 4.3 $\mu$m is not reproduced.  Furthermore, the longitudinal homogeneity in the \ce{CO2} abundances resulting from transport-induced quenching is at odds with the larger \ce{CO2} feature at the evening terminator. In the following subsection, we explore whether the introduction of photochemistry leads to an asymmetry in the CO2 distribution.

\subsection{2D Photochemical Results}
\label{subsec:vulcanresults}
\begin{figure*}%
    \centering
    \includegraphics[width=0.8\textwidth]{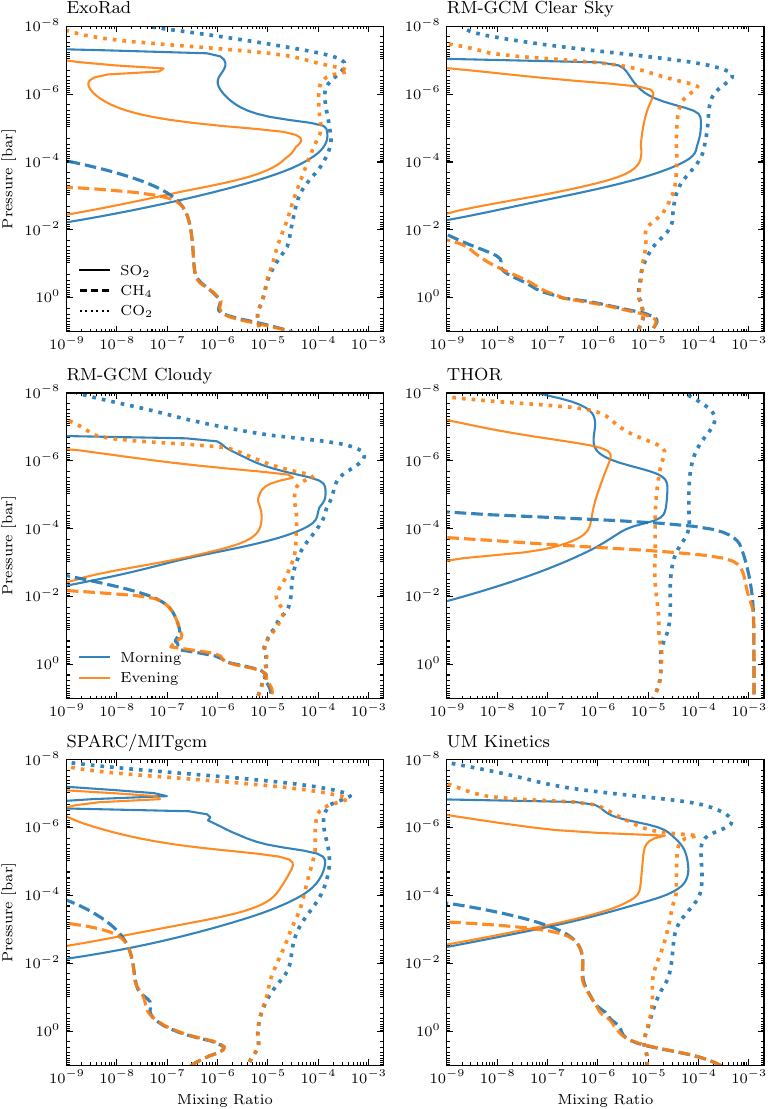}
    \caption{The vertical distribution of \ce{CO2} (dotted), \ce{SO2} (solid), and \ce{CH4} (dashed) at the morning (blue) and evening (orange) terminators for each GCM. \ce{CO2} and \ce{SO2} show moderately higher abundances at the morning terminator, while \ce{CH4} is homogenized by atmospheric transport at pressures above $~1$~mbar and destroyed at lower pressures. }
    \label{fig:co2-so2-limbs}
\end{figure*}

\begin{figure*}
   \centering
   \includegraphics[width=\textwidth]{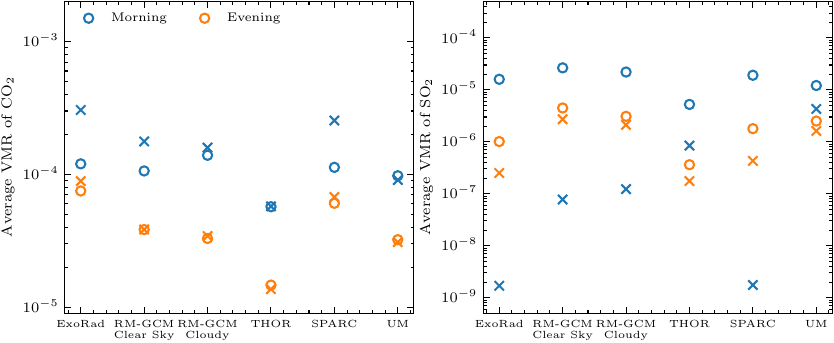}
   \caption{The abundance of \ce{CO2} (left) and \ce{SO2} (right) from the VULCAN 2D model, averaged over 1 to 10$^{-3}$ mbar at morning and evening terminators. Circles are results including zonal transport, while crosses exclude zonal transport (only vertical mixing). While horizontal transport only has a small impact on \ce{CO2} abundances, it strongly enhances the abundances of \ce{SO2} at the morning terminator, due to nightside buildup of \ce{SO2} from photochemical products being transported by the equatorial jet.}
   \label{fig:ave_co2_so2} 
\end{figure*}

\begin{figure}
    \centering
    \includegraphics{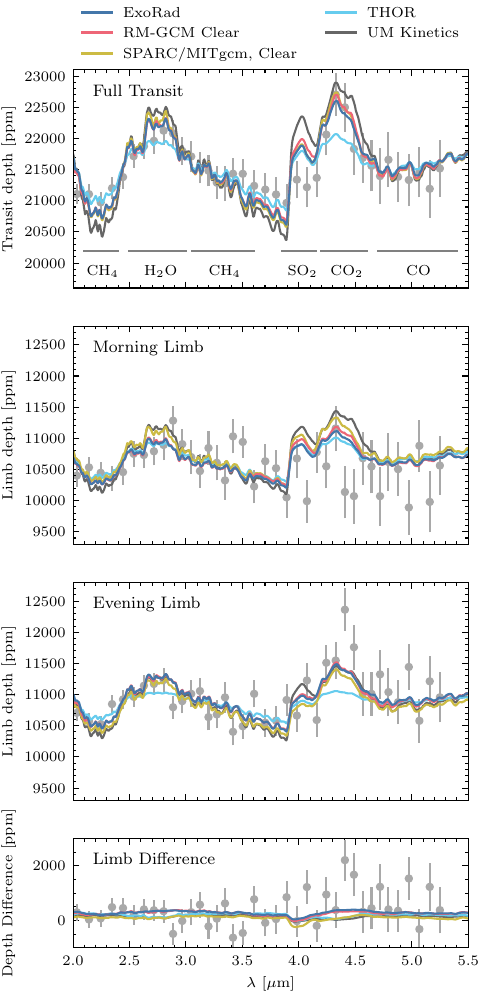}
    \caption{Transmission spectra for the photochemistry (\BOne) post-processing with the observational data in gray. This post-processing applies the 2D photochemical output to all latitudes. Compared to the equilibrium chemistry spectra (Fig.~\ref{fig:trans_clear}), these spectra have strong \ce{SO2} features and no \ce{CH_4} features.}
    \label{fig:trans_photochem_b1}
\end{figure}

\begin{figure}
    \centering
    \includegraphics{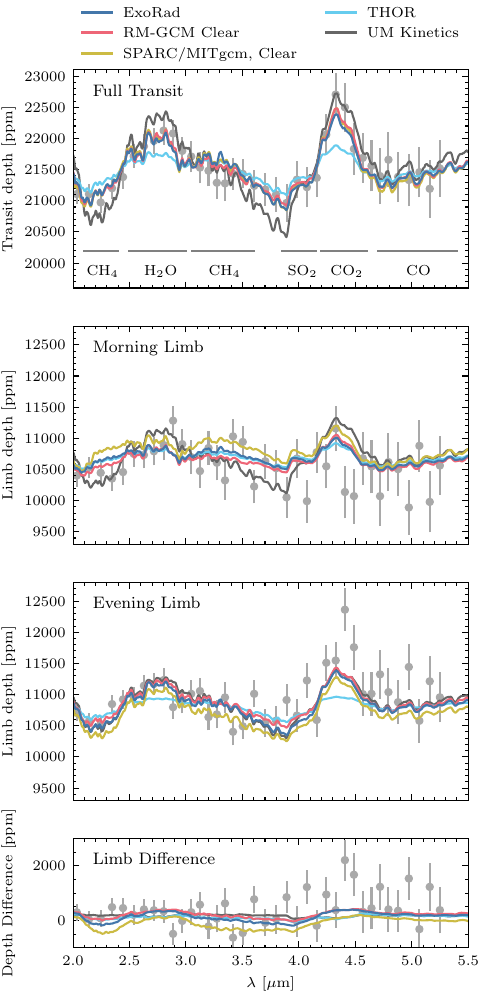}
    \caption{Transmission spectra for the photochemistry (\BTwo) post-processing with the observational data in gray.    This post-processing applies the 2D photochemical output within 20\degree of the equator. Compared to the equilibrium chemistry spectra (Fig.~\ref{fig:trans_clear}), these spectra have moderate \ce{SO2} and \ce{CH_4} features.}
    \label{fig:trans_photochem_b2}
\end{figure}
\subsubsection{Abundances and Vertical Profiles}
As discussed in the previous subsection, atmospheric circulation effectively homogenizes the limb asymmetries present in thermochemical equilibrium. We now examine our 2D photochemical results to assess how photochemistry contributes to limb asymmetries.  Figure \ref{fig:co2-so2-limbs} shows the vertical distributions of \ce{CO2}, \ce{SO2}, and \ce{CH4} at the morning and evening terminators, as computed by 2D VULCAN. Overall, \ce{SO2} and \ce{CO2} exhibit moderate asymmetries in the probed region, while \ce{CH4} shows no morning-evening abundance differences for $p\gtrsim$1~mbar. Above this level, \ce{CH4} is efficiently destroyed by stellar UV radiation, in contrast to the relatively higher abundances in the absence of photochemistry, as seen in Figure \ref{fig:limb_chem}.

The horizontal variation of \ce{CO2} is driven by thermochemistry, while that of \ce{SO2} is driven by photochemistry. \ce{CO2} maintains close to chemical equilibrium throughout the atmosphere up to 10$^{-6}$ bar (see Fig. \ref{fig:co2-so2-limbs}), with higher abundances on the morning terminator where temperatures approach the optimal conditions for \ce{CO2} stability \citep{Moses2013}. Across all models, the limb asymmetry in \ce{CO2} abundance is moderate, differing only by a factor of a few. Generally, differences in \ce{CO2} abundances correlate with temperature differences, i.e. the cloudy RM-GCM and the THOR model exhibit the largest \ce{CO2} limb asymmetries, while the ExoRad model shows the smallest. On the other hand, \ce{SO2} is produced through photochemical processes. While it might initially be expected that the cooler morning terminator would yield less \ce{SO2} due to reduced oxidizing power at lower temperatures \citep{Tsai2023a}, Figure \ref{fig:co2-so2-limbs} shows the opposite. The \ce{SO2} abundances are about an order of magnitude higher at the morning terminator compared to the evening across all models. 

\subsubsection{Horizontal Transport Effects}
The underlying driver for this \ce{SO2} asymmetry becomes apparent in Figure \ref{fig:ave_co2_so2}. While the cooler morning terminator would indeed produce less \ce{SO2} in the absence of horizontal transport, horizontal transport alters the distribution of \ce{SO2}. With horizontal transport, \ce{SO2} can be transported to the nightside and build up to higher abundances in the upper atmosphere since the photodissociation of \ce{SO2} is absent \citep{Tsai2023b, baeyens2024}. This nightside buildup of \ce{SO2} is subsequently transported to the morning terminator, resulting in the \ce{SO2} limb asymmetry. As summarized in Figure \ref{fig:ave_co2_so2}, while \ce{CO2} exhibits little limb asymmetry, \ce{SO2} has enhanced morning abundances induced by horizontal transport from the nightside.

\subsubsection{Impact on Transmission Spectra}
To examine the impact of photochemistry on the limb spectra, an assumption has to be made about how the abundances from the 2D model are mapped back onto the 3D grid used for post-processing (see also Section \ref{subsec:spectra_description}). We explore two cases that should be understood as limiting cases bounding the possible impact of photochemistry on the spectra: \BOne\  (Fig. \ref{fig:trans_photochem_b1}), where the abundances from 2D VULCAN are assumed to extend to all latitudes, and \BTwo\  (Fig. \ref{fig:trans_photochem_b2}), where 2D VULCAN abundances are only used in the equatorial region and equilibrium chemistry is assumed for higher latitudes. The most realistic spectrum can be expected to lie somewhere in between these cases, with spectral features of specific species being closer to either the \BOne\  or the \BTwo\ post-processing depending on their chemical behavior. For example, for \ce{CH4}, which is efficiently destroyed by photochemistry, the true feature amplitude would be expected to be closer to the \BOne\ ~post-processing. For \ce{CO2}, which remains relatively close to thermochemical equilibrium around the globe, the true feature amplitude may be closer to the \BTwo\  post-processing which includes higher \ce{CO2} abundance at the colder mid-latitudes. For \ce{SO2}, where horizontal transport from the nightside through the jet plays a key role, the \BTwo\ post-processing might be more accurate.

The two largest effects of 2D photochemistry on our model spectra are a decrease in the \ce{CH4} feature between 3.0 and 3.5~$\mu$m, and the appearance of the $\rm SO_2$ feature blueward of 4 microns. Changes to the \ce{CO2} feature are small, and thus the inclusion of photochemistry has the potential to improve agreement with observations relative to clear sky models (Section \ref{sec:temperaturestructure}) by altering the contrast between the 4.3 $\mu$m \ce{CO2} peak and the blueward 3.9 $\mu$m window .  The effect of photochemistry on \ce{CH4} can be most clearly seen by comparing the SPARC/MITgcm equilibrium chemistry, \BOne, and \BTwo\   post-processing. In the equilibrium chemistry spectrum (Fig.~\ref{fig:trans_clear}), the morning-limb methane feature is clearly present---in fact, it is the strongest of all models considered, as the SPARC/MITgcm morning limb is the coldest at the top of the atmosphere of all models considered (Fig.~\ref{fig:limb_pt}). In the \BTwo\   post-processing case, the strength of this feature decreases modestly (Fig.~\ref{fig:trans_photochem_b2}), and it disappears entirely in the \BOne\ photochemistry case (Fig.~\ref{fig:trans_photochem_b1}). That is, as we extend the abundances from the 2D photochemistry model to the full globe, the spectral signature of $\rm CH_4$ correspondingly decreases as atmospheric transport quenches the molecule to a lower abundance at the morning terminator. Further, $\rm CH_4$ is photodissociated at low pressures. Thus, the spectral differences between morning and evening terminator in the $\rm CH_4$ bands disappear. A similar but somewhat less pronounced effect can be seen in the RM-GCM and ExoRad models.

The effect of photochemistry on $\rm SO_2$ is most prominently seen in the UM Kinetics model, but can also be seen in the other models. In the models without photochemistry, the spectra show no sign of $\rm SO_2$ (Fig.~\ref{fig:trans_clear}), owing to low equilibrium $\rm SO_2$ abundances (Fig.~\ref{fig:limb_chem}) and, in the case of the UM Kinetics model, the non-inclusion of \ce{SO2} in the kinetics network due to its expected low abundance when only considering thermochemistry. Once photochemistry is taken into account, however, the SO2 feature becomes more prominent, with the \BOne\ post-processing method yielding the largest signal (Figs.~\ref{fig:trans_photochem_b1}--\ref{fig:trans_photochem_b2}). As noted above, transport brings $\rm SO_2$ in higher abundance to the morning limb, yielding stronger morning $\rm SO_2$ absorption than evening. The signal-to-noise in the current morning and evening spectra of WASP-39b is too low to unambiguously detect \ce{SO2} \citep{espinoza2024inhomogeneous,TadaEtAl2025WASP-39bLimbAsymmetries}. However, detecting differences in the \ce{SO2} feature may be feasible in the future by combining multiple transits.

\section{Condensate Clouds}
\label{sec:clouds}

In this section we examine the effect of clouds on limb asymmetries using the two complementary approaches: by post-processing the ExoRad GCM simulations with the IWF-Graz cloud model (described in Section \ref{subsec:IWF_cloud_description}) and through the inclusion of radiatively-active clouds in the RM-GCM (described in Section \ref{subsec:RMGCM_cloud_description}).  A further summary of these models and useful quantities is given in  Table~\ref{tab:cloud_definitions} in Appendix \ref{sec:cloud_properties}.

\begin{figure*}
    \centering
    \includegraphics[width=0.45\linewidth]{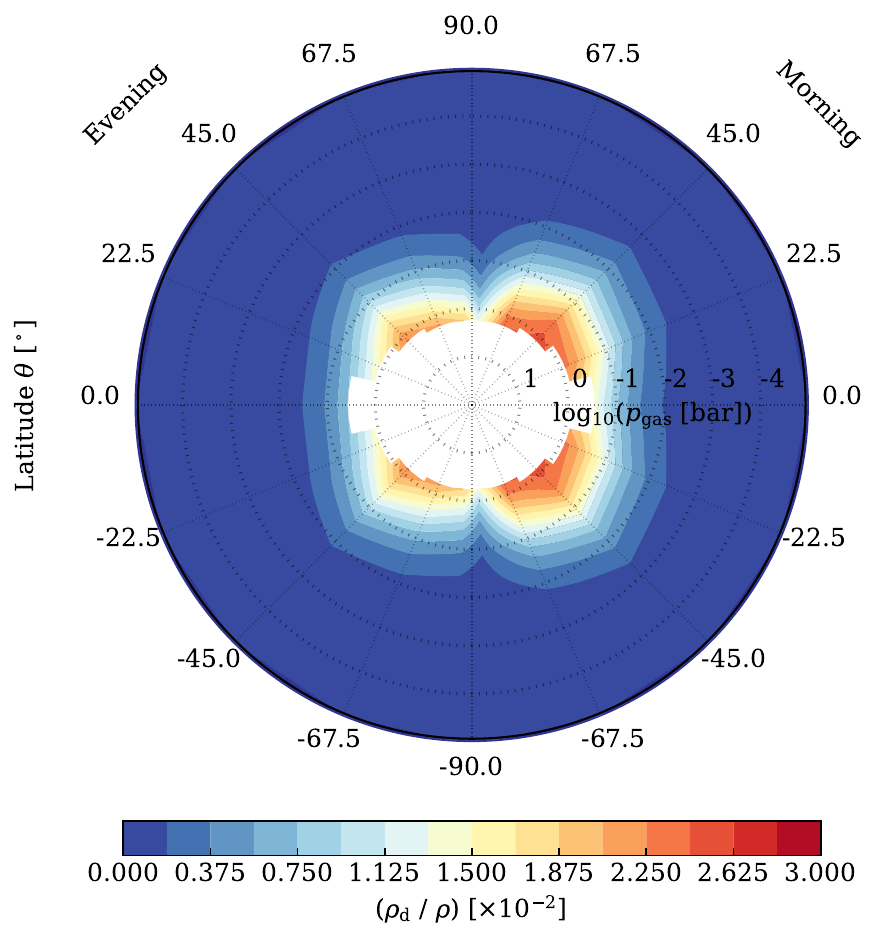}
     \includegraphics[width=0.45\linewidth]{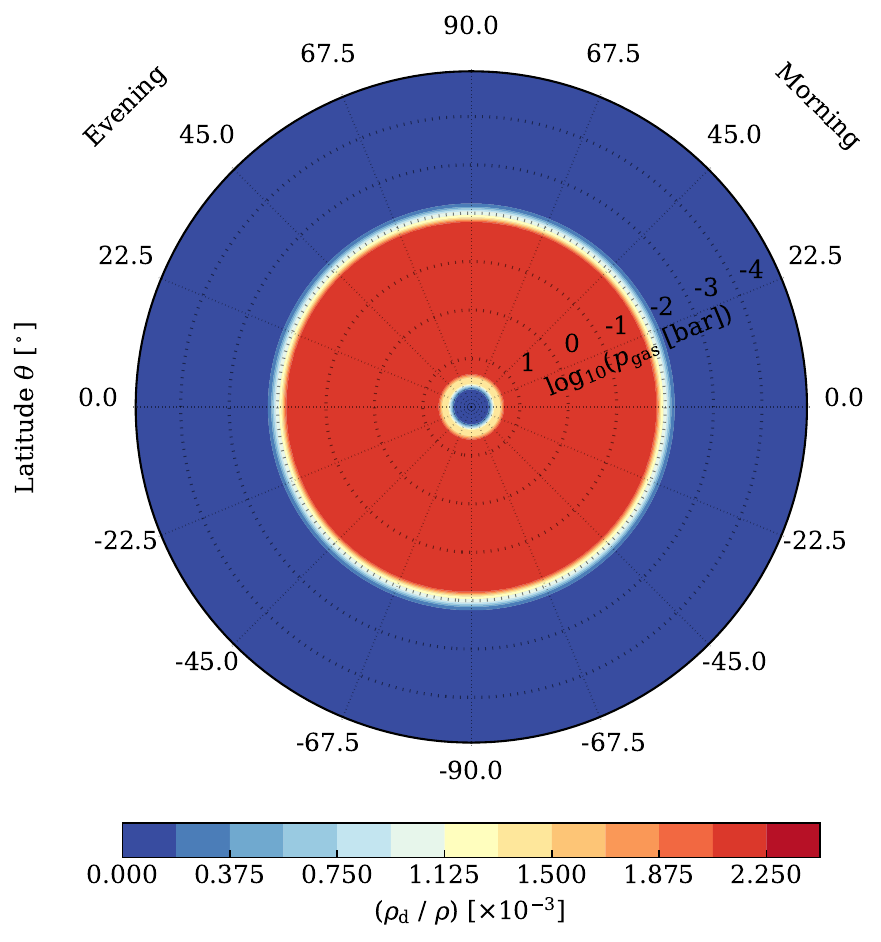}
    \caption{%
     The local cloud mass fraction or dust-to-gas mass ratio $\rho_{\rm d}/ \rho$  for the atmosphere of WASP-39b in the terminator plane (evening: left hand side of each 2D slice, morning: right hand side of each 2D slice). The left panel shows  the ExoRad/IWF~Graz cloud model and the right panel the RM-GCM cloud model, respectively. The white area in the center of the left panel is outside the modelled domain of the IWF~Graz cloud model. Generally, the cloud mass fraction in the ExoRad/IWF~Graz model is roughly a factor 10 higher than in the RM-GCM. The ExoRad-IWF~Graz model also predicts noticeable variations with latitude and between morning and evening terminator, while the RM-GCM exhibits a more homogeneous cloud mass fraction. Please note the change in scale between the two panels.}
    \label{fig:rho_d}
\end{figure*}

\begin{figure*}
    \centering
    \includegraphics[width=0.49\linewidth]{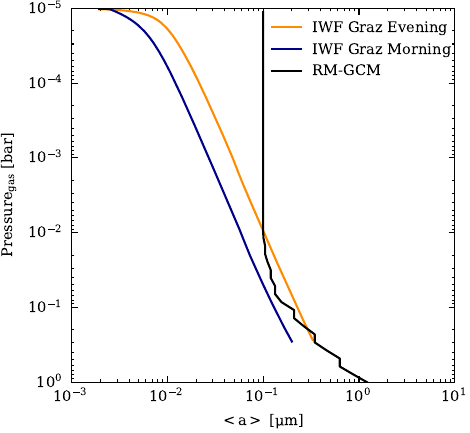}
    \includegraphics[width=0.49\linewidth]{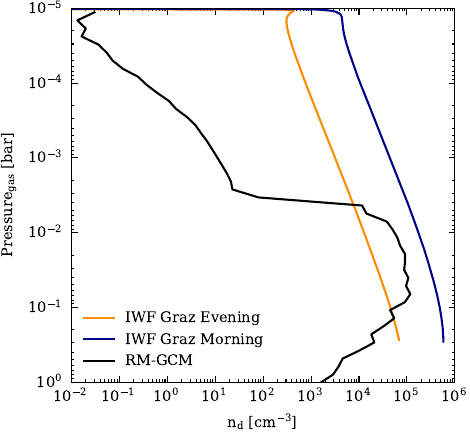}
    \caption{ Mean particle size $\langle a \rangle $ (left) and cloud particle number density $n_d$ (right) for the equatorial morning  (blue) and evening terminator (orange) for the ExoRad/IWF Graz model and for the cloudy RM-GCM (black). The ExoRad/IWF Graz model predicts larger particle sizes and lower number densities at the evening terminator than at the morning terminator. In contrast, the cloudy RM-GCM assumes the same prescribed vertical particle size profile at all locations. }
    \label{fig:mean_a_vsp}
\end{figure*}

\begin{figure*}
    \centering
    \includegraphics[width=0.49\linewidth]{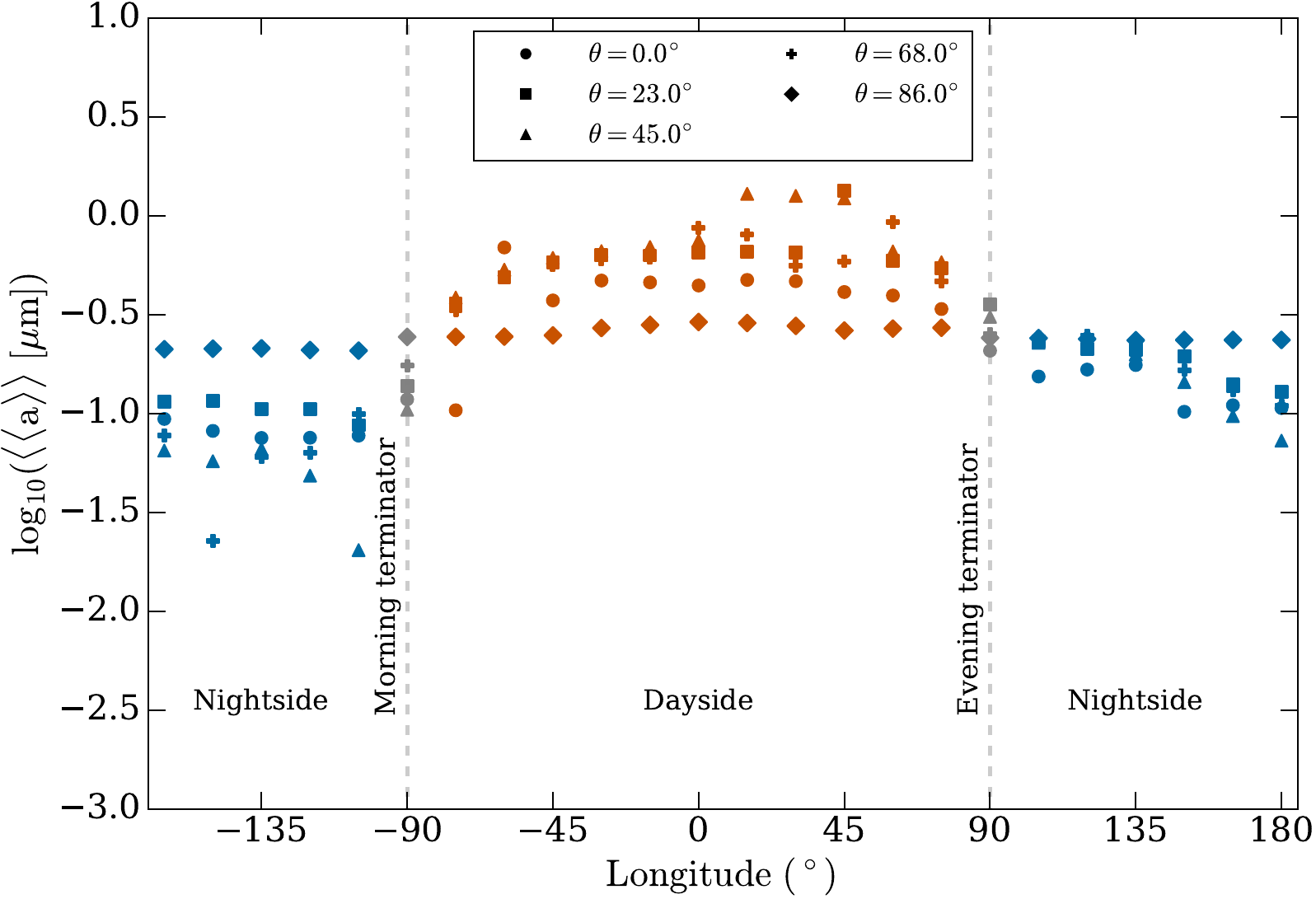}
       \includegraphics[width=0.49\linewidth]{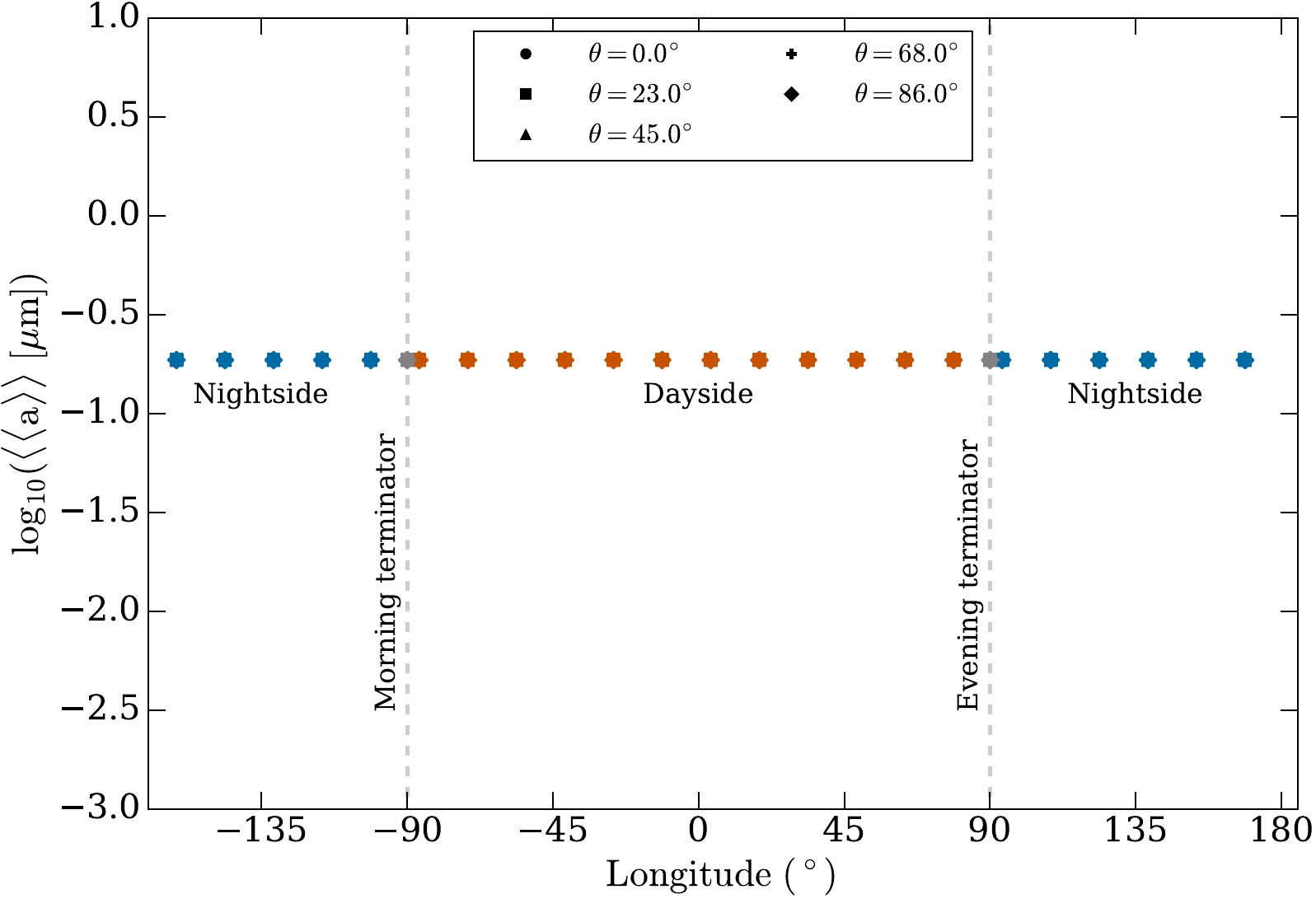}
    \caption{Column integrated (from $p=10^{-1}$~bar to $p=10^{-5}$~bar), number-density-weighted average particle size, $\log_{10}\left( \langle \langle a \rangle \rangle \right)$ at different locations in longitude and for five representative latitudes for the ExoRad/IWF~Graz model (left) and the cloudy RM-GCM (right), respectively. The ExoRad/IWF~Graz model predicts substantial horizontal variations in the average particle size, with larger particles on the dayside than on the nightside, while the RM-GCM assumes a horizontally uniform particle size.}
    \label{fig:mean_a}
\end{figure*}

\begin{figure}
    \centering
    \includegraphics[width=0.8\columnwidth]{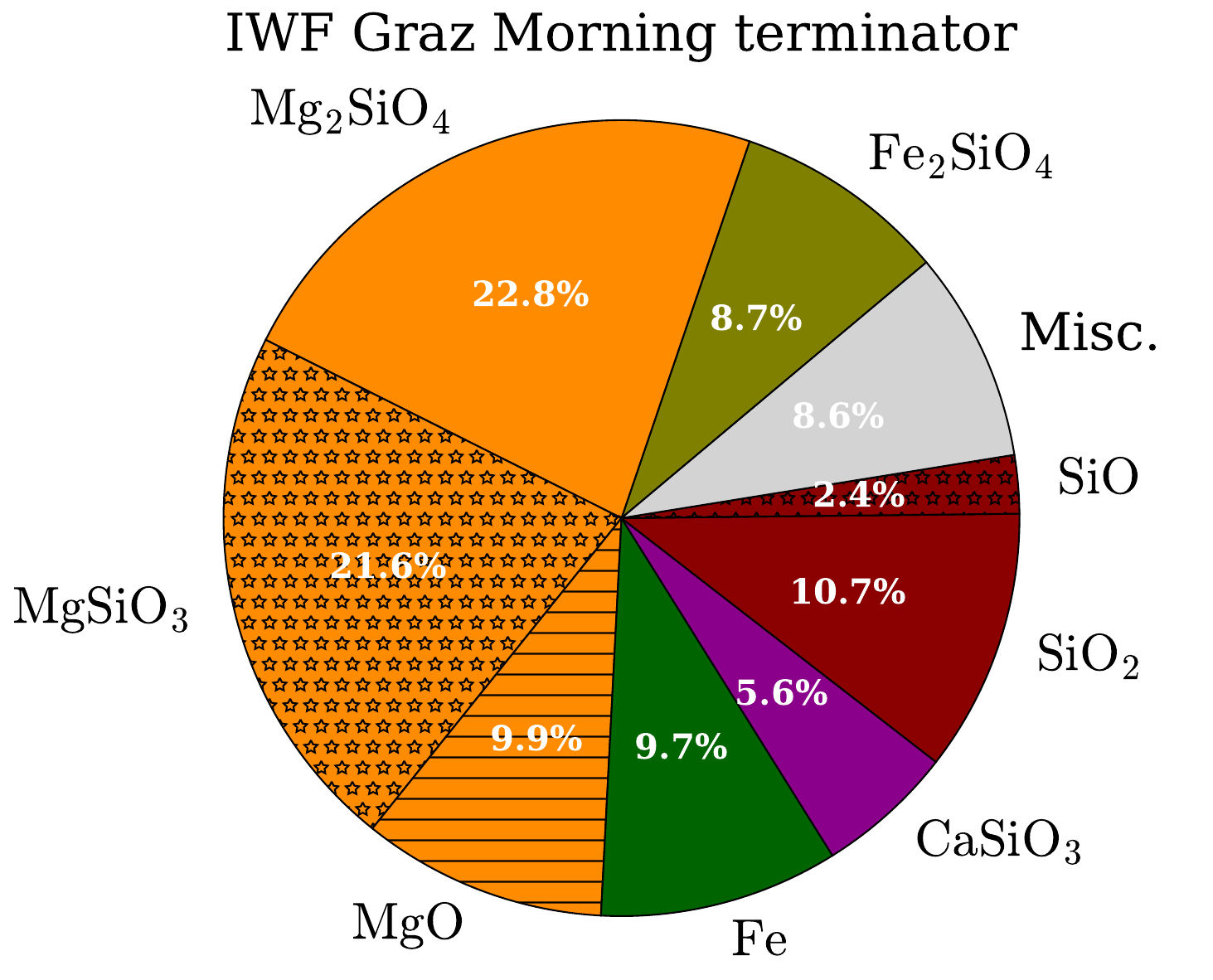}\\
    \vspace{5pt}
        \includegraphics[width=0.8\columnwidth]
        {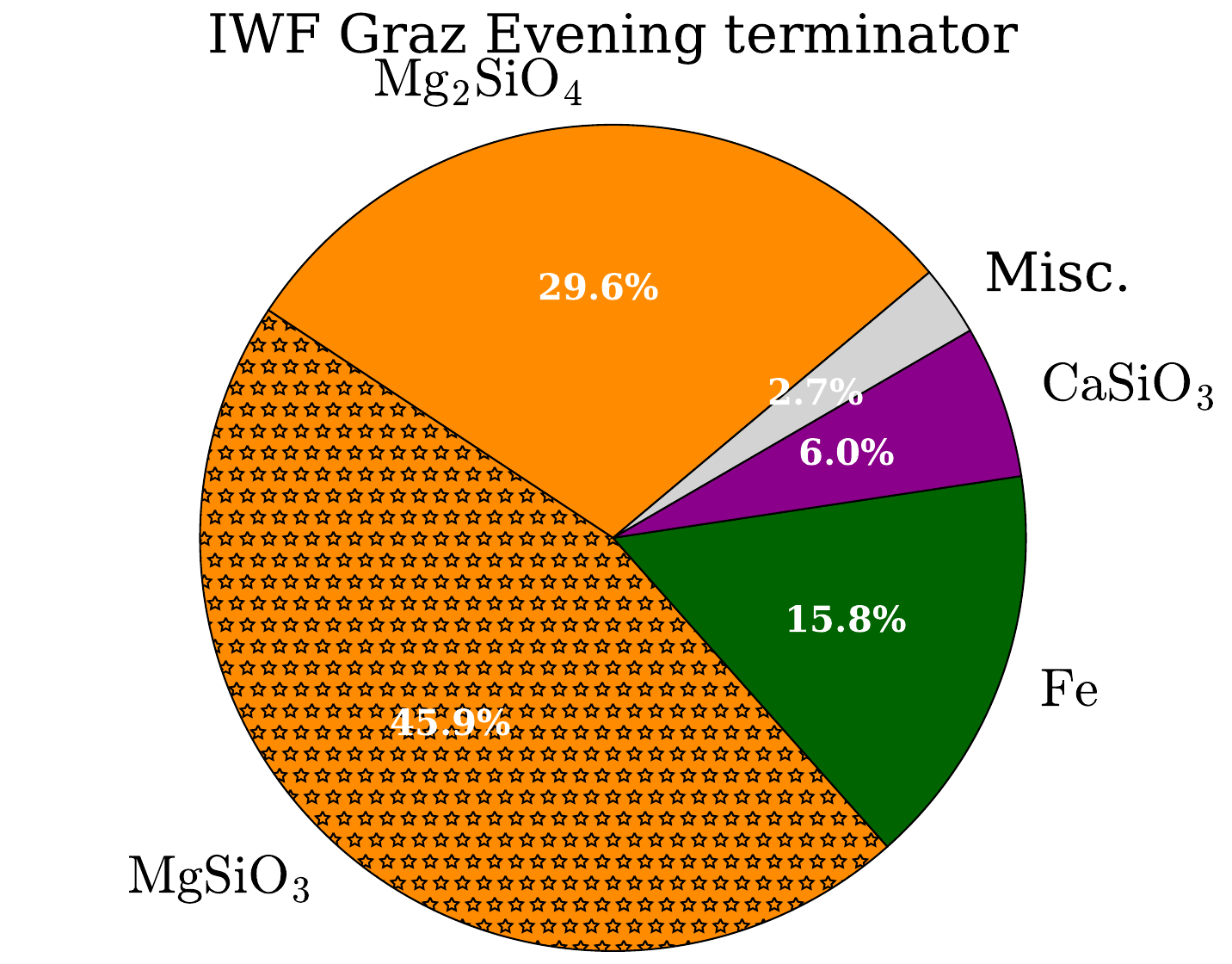}\\
        \vspace{5 pt}
        \includegraphics[width=0.8\columnwidth]{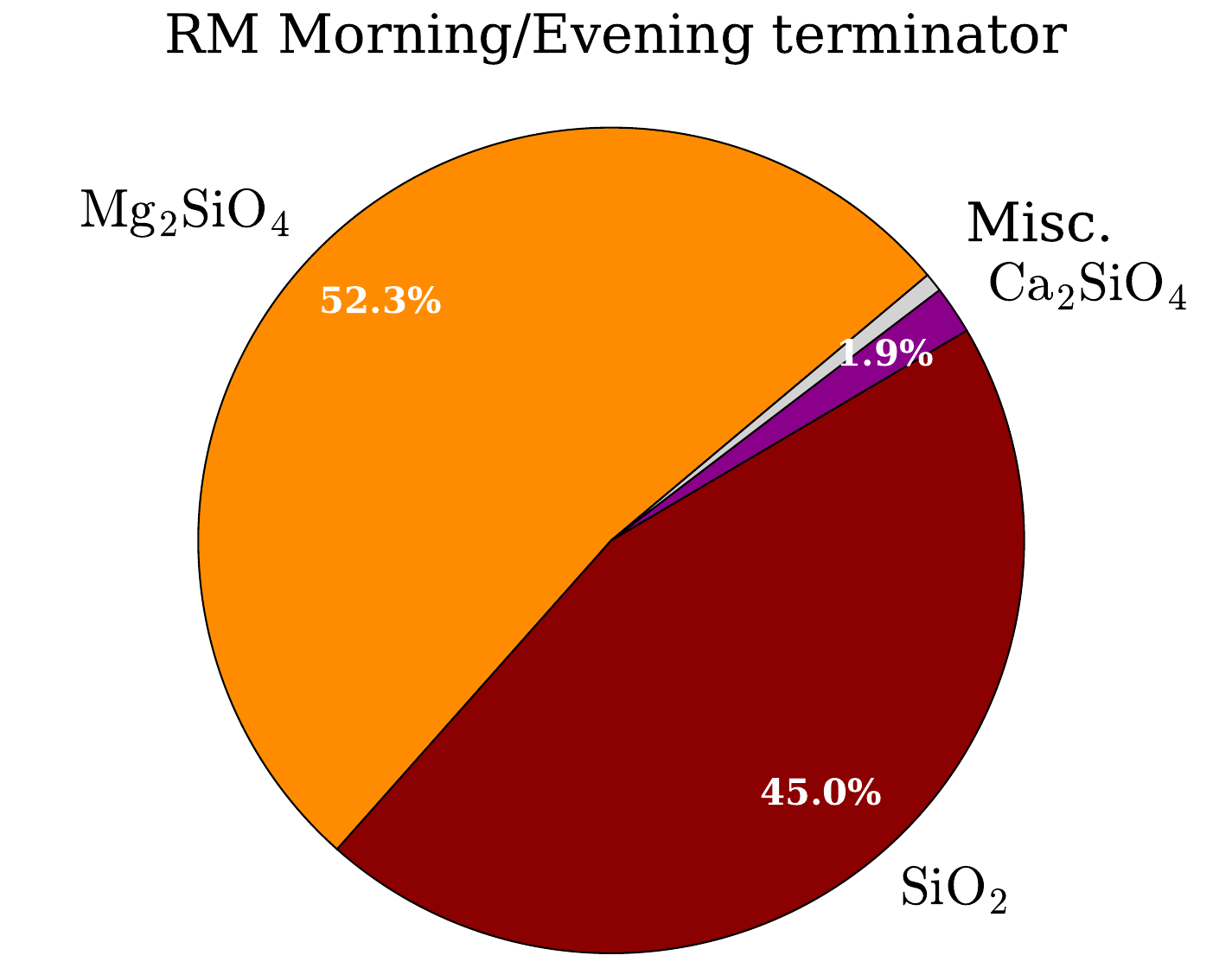}
    \caption{Composition of the clouds at the morning and evening terminators at $p_{\rm gas}=10^{-2}$~bar in volume fractions, $V_s/V_{tot}$, for the equatorial morning terminator (left) and evening terminator (middle) of WASP-39b from the ExoRad/IWF~Graz model and both terminators from RM-GCM (right). The ExoRad/IWF~Graz model produces mixed-material particles that differ between both limbs because the evening terminator is substantially hotter. The RM-GCM instead assumes condensates of homogeneous compositions, with relative abundances determined by the assumed condensation temperatures and gaseous abundances, which are similar across the terminator. 
    }
    \label{fig:cloud_comp}
\end{figure}

\subsection{Cloud Mass Load and Horizontal Distribution}
Comparing the cloud properties arising from the two 3D cloud models, we find that the ExoRad/IWF~Graz model
predicts stronger horizontal variations and a higher cloud mass fraction. Figure~\ref{fig:rho_d} shows the cloud mass fraction for a slice across the terminator, as it would appear to an observer during transit. 
It is immediately clear that the RM-GCM model, with parameterized clouds in the GCM, produces a very homogeneous distribution and density of clouds at both limbs of the terminator with a significant jump in density at $10^{-2}\,{\rm bar}$, which is unsurprising given the combination of the assumed vertical truncation and all major cloud species first condensing in the (horizontally homogeneous) deep atmosphere. Comparatively, the ExoRad/IWF~Graz model produces a smoother increase in total cloud mass fraction with increasing pressure and shows more variation with latitude and for each limb. There is also a substantially higher cloud mass fraction for the ExoRad/IWF~Graz model than the RM-GCM, by approximately one order of magnitude (see Fig.~\ref{fig:rho_d} caption). This is despite both models assuming a 10 x solar metallicity atmosphere. A likely cause of this difference is that the RM-GCM only allows 10\% of available cloud-forming mass to condense. While all simplified cloud models have to make a somewhat arbitrary choice about which fraction of the available cloud-forming mass condenses (a question only detailed microphysical modeling or observational constraints can answer), the value of 10\% chosen here was at least partially motivated by numerical stability. Further, we note that in the RM-GCM model, the cloud mass fraction only depends on whether the temperature is above or below the condensation temperature and is not sensitive to the exact value of the local temperature. As the limb temperature differences in the RM-GCM model are not large enough to put one limb above and one limb below the condensation temperature (except for KCl, a minor cloud component), the clouds are homogeneous across the terminators. Conversely, the cloud mass load from the microphysical cloud model used in the ExoRad/IWF~Graz model is more sensitive to moderate temperature differences of a few 100~K and as a result exhibits significantly more cloudiness at mid-latitudes on the morning terminator \citep[see also][]{Carone2023}.

\subsection{Particle Sizes and Number Densities}
These differences in cloud models also extend to particle sizes and number densities. Figure~\ref{fig:mean_a_vsp} displays the average particle size $\langle a \rangle$ and the number density of cloud particles $n_{\rm d}$ for the equatorial points of the morning and evening terminators. These selected points are qualitatively representative of the differences between the models at all latitudes. We note that in the RM-GCM, the particle size follows a prescribed vertical profile at all longitudes and latitudes, and thus there are no horizontal differences in the particle size. The two models display comparable particle sizes and number densities of cloud particles for pressures greater than a few mbar. They differ substantially for lower pressures, where the ExoRad/IWF~Graz model predicts smaller particles and a much less steep decline in number density than the RM-GCM.  
The ExoRad/IWF~Graz model also predicts mean particle sizes to vary substantially between day- and night-side in contrast to the RM-GCMs horizontally-uniform mean particle size (Fig.~\ref{fig:mean_a}). On the dayside, the ExoRad/IWF~Graz model generally has larger average sizes, as the higher temperatures restrict the amount of nucleation occurring in the upper atmosphere, and consequently more bulk growth occurs to the limited number of cloud particles \citep[see][for a more detailed discussion]{Helling2023_Grid}. In the RM-GCM, the column-integrated mean radius remains at $\sim 0.2\,\mu{\rm m}$ globally due to the assumed particle size profile in the model.  Despite these differences, the particle sizes assumed in the RM-GCM model are qualitatively similar to those predicted by the IWF/Graz model at the terminator.  

\subsection{Cloud Material Composition}
We consider the material compositions of the clouds, where we focus on the $p_{\rm gas} = 10^{-2}\,{\rm bar}$ pressure level for the equatorial morning and evening terminators (Fig.~\ref{fig:cloud_comp}).
While the material mix is ``simpler'' for the RM-GCM cloud model compared to the morning terminator of the ExoRad/IWF~Graz model, both cloud models agree that a large part of the morning cloud at $p=10^{-2}$ should consist of olivines  and enstatite ((Fe/Mg)$_2$SiO$_4$[s] and MgSiO$_3$[s]) that combined make up about 50\% of the cloud material. Larger discrepancies in cloud composition arise with respect to silica clouds composed of SiO$_2$[s], which only make up 11\% of the cloud volume in the kinetic cloud formation ExoRad/IWF~Graz model at the morning terminator, and are completely missing at the warmer evening terminator. In the RM-GCM cloud model, which assumes all cloud species condense when the gas phase is saturated, SiO$_2$[s] contributes more than 40\% at both limbs and is thus the second most abundant cloud species.
Furthermore, the ExoRad/IWF-Graz model also includes iron-bearing condensates and assumes that iron is not already rained out due to condensation in the deep atmosphere. As a result, significant fractions of Fe[s] are found at the evening terminator, and both Fe[s] and \ce{Fe2SiO4}[s] at the morning terminator. Iron-bearing species are highly absorbing and have been shown to have a strong impact on transmission spectra in mixed-grain particles even when they make up less than 1~\% of particle volume \citep{Kiefer2024}. The absence of \ce{Fe2SiO4}[s] at the evening terminator is explained by the hotter temperatures of the model. In contrast, iron does not take part in cloud formation in the RM-GCM.
Another difference between the ExoRad/IWF~Graz and RM-GCM cloud model is that the former shows notable composition differences between the limbs. The warmer evening terminator leads to a reduction in both the number of different species and the dust-to-gas mass ratio in the ExoRad/IWF~Graz model.

\subsection{Impact of Clouds on Temperature Structure}
\begin{figure*}
    \centering
    \includegraphics{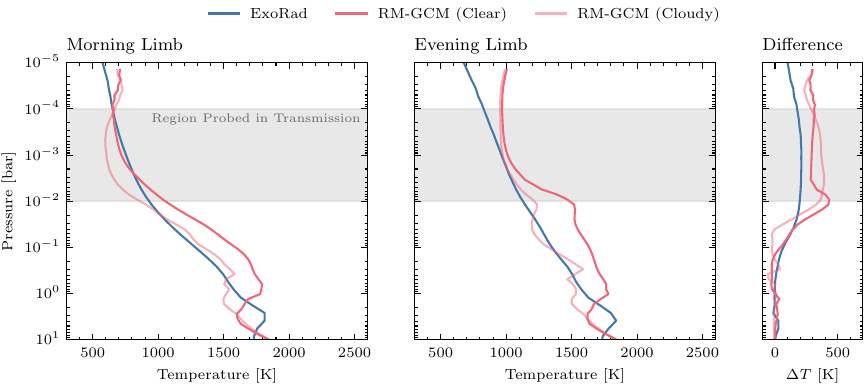}
    \caption{As Fig. \ref{fig:limb_pt} but for the models that include clouds. For the RM-GCM, which contains cloud radiative feedback, the clear-atmosphere temperature profile is shown for comparison. The ExoRad/IWF Graz model post-processed clouds onto the temperature structure from the clear-atmosphere GCM, thus only one profile is shown.}
    \label{fig:limb_pt_cloud}
\end{figure*}

\begin{figure*}
    \centering
    \includegraphics{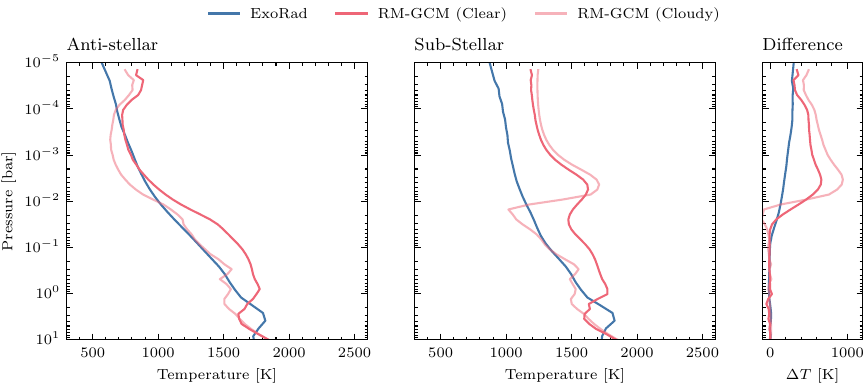}
    \caption{As Fig. \ref{fig:daynight_pt} but for the models that include clouds. For the RM-GCM, which contains cloud radiative feedback, the clear-atmosphere temperature profile is shown for comparison. The ExoRad/IWF Graz model post-processed clouds onto the temperature structure from the clear-atmosphere GCM, thus only one profile is shown.}
    \label{fig:daynight_pt_cloud}
\end{figure*} 

Examining the impact of clouds on the temperature structure with the RM-GCM, we find that cloud feedback leads to cooler temperatures at both terminators (Fig. \ref{fig:limb_pt_cloud}). In the cloudy RM-GCM, clouds are nearly globally present for $p>3\times10^{-3}$~bar, and effectively absent at lower pressures as a consequence of the assumed vertical truncation (see also Figs. \ref{fig:rho_d} and \ref{fig:mean_a_vsp}). The presence of this cloud deck results in the rapid absorption and reflection of starlight by $p\sim10^{-2}$~bar, resulting in a strong ($\approx 700$~K) thermal inversion on the dayside (Fig. \ref{fig:daynight_pt_cloud}). Overall, the rapid attenuation of starlight enhances day-night temperature differences at this pressure and lower pressures. At higher pressures, the opposite effect is produced. Increased opacity from clouds in the thermal channels (and therefore longer radiative timescales) combined with the lack of starlight to drive asymmetries produces a significantly more homogeneous intermediate and deep atmosphere. Additionally, scattering by dayside clouds increases the planetary albedo, cooling the planet overall. 

At the morning terminator, the combination of these effects results in cooling by about 100~K throughout most of the atmosphere. At the evening terminator, there is less cooling in the region probed by transmission spectra ($<50$~K), yet stronger cooling ($200-400$~K) at $p>10^{-2}$~bar.

\subsection{Impact on Transmission Spectra}
\begin{figure}
    \centering
    \includegraphics[width=\columnwidth]{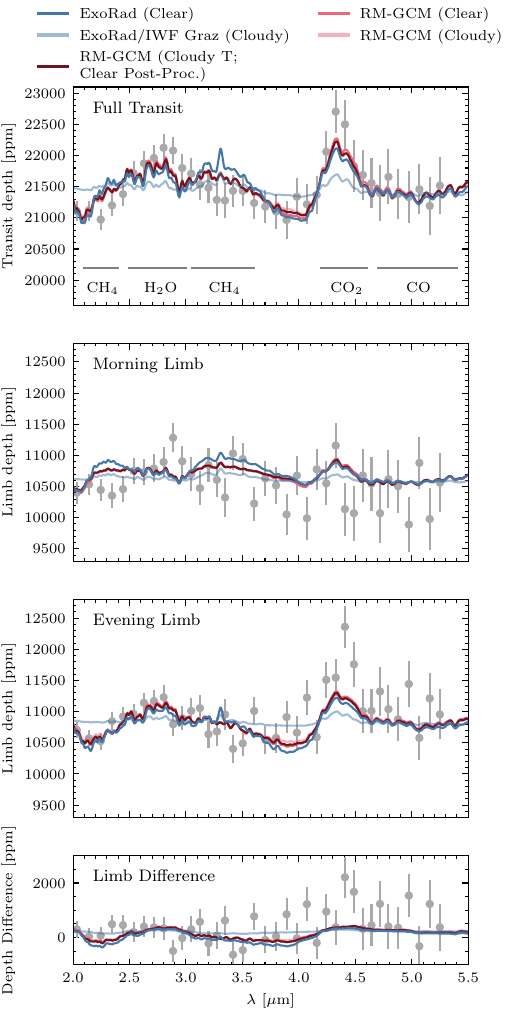}
    \caption{Transmission spectra for the cloudy simulations with the observational data in gray, with the corresponding clear spectra shown for comparison. To distinguish the impact of temperature changes due to cloud radiative feedback from the direct effect of cloud opacity on the transmission spectrum in the RM-GCM, the crimson line shows the transmission spectrum of the cloudy RM-GCM post-processed without cloud opacity. The full transmission spectra are in the top panel with the individual limbs shown in the middle panels. The difference between the limbs is shown in the bottom panel. In the ExoRad/IWF~Graz model, clouds strongly decrease the strength of spectral features at both limbs. The RM-GCM exhibits much smaller changes, and some of the changes in the RM-GCM are driven by temperature changes due to cloud radiative feedback rather than cloud opacity.}
    \label{fig:trans_clouds}
\end{figure}

In Figure \ref{fig:trans_clouds}, we compare the cloudy spectra for both models to each other, to the cloud-free models, and to the observations. The cloudy spectra are post-processed with the same framework as the clear-atmosphere models (see Section~\ref{subsec:spectra_description}). Thus, the post-processing differs from the 1D framework used for the cloudy models in \citetalias{espinoza2024inhomogeneous} (see Appendix~\ref{sec:spec_differences}). 

Clouds have a strong impact on the spectrum with the ExoRad/IWF Graz model, with differences to the clear atmosphere spectrum of up to $\approx 500$~ppm in some spectral regions. For this model, all spectral features are strongly muted, both in the evening and the morning spectrum, though the centers of the CH$_4$ and CO$_2$ features remain visible. Including clouds overall improves the match to observations for the morning limb ($\chi^2_{\rm red}$ of 1.14 compared to 1.52 for the cloud-free model, see Appendix \ref{sec:chisquared}), primarily by decreasing the amplitude of the CH$_4$ features. However, clouds simultaneously worsen the match for the evening limb and full transit spectrum by muting spectral features.

In the RM-GCM, the differences between the cloudy and the clear spectra are much smaller ($\approx 100$~ppm) and are partially driven by temperature changes rather than cloud opacity directly. Cloud opacity only contributes notably to the spectral shape at the evening terminator in the spectral regions between 2 to 2.5~$\mu$m and 3.7 to 4 $\mu$m, which are spectral windows probing deeper layers of the atmosphere. In these spectral regions, the impact of temperature changes and cloud opacity on the spectrum are comparable and add up. At other wavelengths at the evening terminator, only the temperature change influences the spectrum, and the overall effect on the spectrum is even smaller ($\lessapprox 50$~ppm). At the morning terminator, the difference between the cloudy and clear RM-GCM is negligible. This is not due to smaller temperature changes---in fact, the temperature change due to cloud feedback is larger at the morning terminator than at the evening terminator in the region probed in transmission. However, with the assumption of equilibrium chemistry used in the calculation, it may be the case that the decrease in feature size due to colder temperatures is offset by the increase in methane abundance at the colder morning terminator. Due to the overall small effects of clouds in this model, the match to the observations only marginally changes.

The stark difference between the ExoRad/IWF~Graz model and the RM-GCM in terms of the impact of clouds on the transmission spectrum can be explained by three factors. The first is the inclusion of iron-containing cloud species in the ExoRad/IWF~Graz model, which has been demonstrated to strongly increase the opacity of mixed-grain clouds \citep{Chubb2024,Kiefer2024}. Even volume fractions of Fe[s] below 1\% in mixed grains can substantially increase the opacity in mixed grains \citep{Kiefer2024}. Furthermore, the inclusion of iron species has been shown to lead to stronger muting of spectral features if opacities are calculated assuming mixed grains using effective medium theory (as in the IWF~Graz model) compared to separate homogeneous particles for each species (as in the RM-GCM) \citep{Kiefer2024}. A recent study using the IWF~Graz model, published while this work was under review, found that excluding iron species from the opacity calculation significantly improved the match to WASP-39b observations \citep{Carone2026}. However, the detailed mechanism for suppressing the formation of iron cloud species remains unclear.

The remaining two factors both stem from modeling choices in the RM-GCM: the assumption that only 10~\% of the available condensable gas mass actually condenses into clouds in the RM-GCM, and the treatment of the cloud thickness as a free parameter. Exploring how changing these assumptions of the cloudy RM-GCM could improve the match to observations remains a task for future work.

Finally, we note that neither of the two cloud models shows any promise for explaining the difference in the amplitude of the CO$_2$ feature between the morning and evening limb. The ExoRad/IWF~Graz model strongly mutes the feature at both limbs, while the change in the CO$_2$ feature due to clouds in the cloudy RM-GCM is marginal. However, as already pointed out in \citet{espinoza2024inhomogeneous}, the clear ExoRad model qualitatively better matches the spectrum at the evening terminator (despite still predicting a too low amplitude of the CO$_2$ feature), while the cloudy ExoRad/IWF~Graz model provides a better match to the morning terminator.  While the ExoRad/IWF~Graz model predicts a lower cloud deck for the evening terminator than for the morning terminator, the data suggest much less cloud coverage at the evening terminator. Thus, future work could explore factors that increase the cloud inhomogeneity to explain the observations, as further discussed in Section \ref{sec:discussion}.

\section{Photochemical Hazes}
\label{sec:hazes_results}

\begin{figure}
    \centering
    \includegraphics[width=\columnwidth]{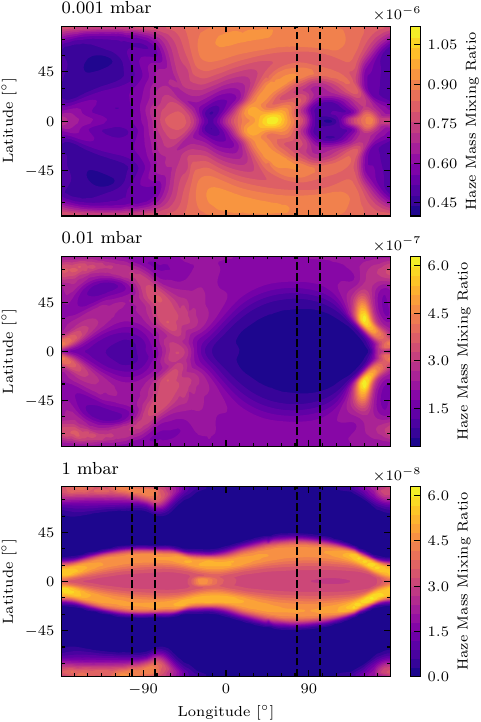}
    \caption{Horizontal distribution of photochemical hazes in the simulation with 30~nm-sized hazes on multiple isobars. The substellar point is located in the center of each panel. The black dashed lines indicate the opening angle along the terminator, computed following \citet{Wardenier2022}. At very low pressures near the peak of the haze production region, hazes are more concentrated at the evening terminator than at the morning terminator (top panel). At pressures typically probed in transmission, however, the haze mass mixing ration is higher at the morning terminator (center and bottom panels). }
    \label{fig:haze_isobar}
\end{figure}

\begin{figure}
    \centering
    \includegraphics[width=\columnwidth]{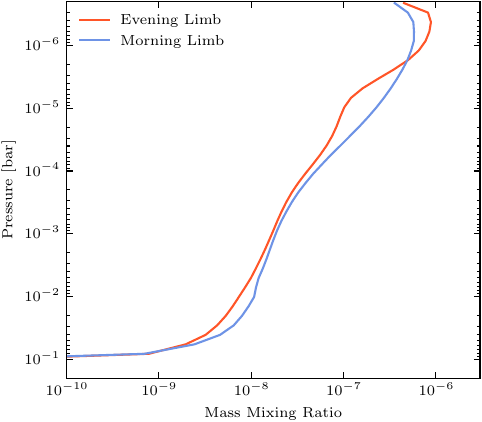}
    \caption{Latitudinally-averaged mass mixing ratio profiles of 30-nm-sized hazes at the morning and evening terminator. At most pressures, the mass mixing ratio is higher at the morning terminator than at the evening terminator.}
    \label{fig:haze_mmr}
\end{figure}

\begin{figure*}
    \centering
    \includegraphics{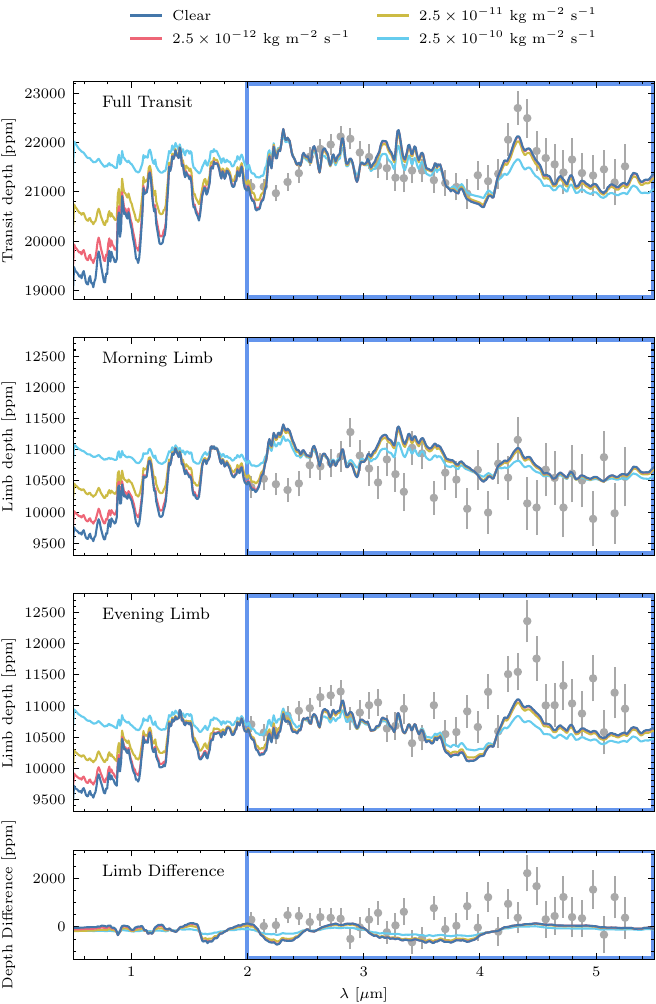}
    \caption{Transmission spectra for the hazy simulations with the observational data in grey. The full transmission spectra are in the top panel with the individual limbs shown in the middle panels. The difference between the limbs is shown in the
    bottom panel. Displayed are spectra from SPARC/MITgcm simulations with photochemical hazes with a particle size of 30~nm  with multiple haze production rates. Generally, hazes have little impact on the spectrum for wavelengths $>2\,\mu$m. At short wavelengths, however, they strongly increase the transit radius and impose a small ($\lessapprox100$~ppm) limb asymmetry resulting in a larger morning terminator. To highlight this effect, the x-axis extends to 0.5~$\mu$m. The blue box highlights the wavelength region shown in the other figures.}
    \label{fig:trans_hazes}
\end{figure*}

As shown in other studies as well \citep{SteinrueckEtAl2021, SteinrueckEtAl2023, MakEtAl2025}, photochemical hazes have a complex 3D distribution. The horizontal distribution of hazes is shown in Figure \ref{fig:haze_isobar} and Figure \ref{fig:haze_mmr} shows the vertical mass mixing ratio profile at the morning and evening terminator. Generally, at pressures below the peak of haze production ($<2 \times 10^{-6}$~bar), the haze mass mixing ratio is higher at the evening terminator, while between $2 \times 10^{-6}$~bar and $10^{-4}$~bar, hazes are concentrated on the night side and near the morning terminator \citep[see also][]{SteinrueckEtAl2021, SteinrueckEtAl2023}. For pressures $\gtrapprox 10^{-4}$~bar, the haze mass mixing ratio is more uniform between evening and morning terminator, with slightly more hazes present at the morning terminator.

Due to the small particle radii expected for photochemical hazes, hazes have a much larger impact on shorter wavelengths than the 2--6 $\mu$m wavelength region we focus on in this study. However, in the case of large haze production rates, hazes can still influence the spectrum in the near-infrared (Fig. \ref{fig:trans_hazes}). In addition, hazes have a much stronger impact on transmission spectra in the short-wavelength region of NIRSpec PRISM and NIRISS SOSS. To demonstrate this impact, we extended the wavelength region shown in Figure \ref{fig:trans_hazes} to shorter wavelengths. Generally, photochemical hazes in our simulations of WASP-39b have two main effects on the transit spectrum: they mute the amplitude of molecular features, and they increase the transit radius more strongly at the morning terminator than at the evening terminator.

Even in the models with large haze production rates, the limb asymmetries signal remains dominated by differences in the chemical composition between morning and evening terminator. For the equilibrium chemistry case, much more methane is present on the cooler morning terminator. Thus, in the clear-atmosphere equilibrium chemistry case, there is a strong negative limb depth difference in methane bands (2-2.5~$\mu$m and 3-4~$\mu$m), while there is a slightly positive limb depth difference outside the methane bands. With increasing haze production rate, molecular features become increasingly muted and thus the spectral variations in the limb asymmetries signal also become less pronounced. 

The effect of the asymmetric haze distribution is most obvious in the wavelength regions $<1.5$~$\mu$m. Because hazes are more concentrated at the morning terminator than at the evening terminator over the pressures probed in transmission, at short wavelengths, the transit radius of the morning terminator becomes larger than the evening terminator in the continuum regions. 

Finally, we point out that while the haze production rate in hot Jupiter atmospheres remains highly unknown, haze production rates as large as the strongest two cases shown in Figure \ref{fig:trans_hazes} ($2.5\times 10^{-11}$ kg~m$^{-2}$~s$^{-1}$ and $2.5\times 10^{-10}$ kg~m$^{-2}$~s$^{-1}$) are considered unlikely for WASP-39b based on shorter-wavelength observations and photochemical models \citep{ArfauxLavvas2022}. However, for some other exoplanets (especially cooler ones), such strong haze production rates remain a possibility.

\section{Combining condensate clouds and disequilibrium chemistry}
\label{sec:photochem+clouds}
\begin{figure}
    \centering
    \includegraphics{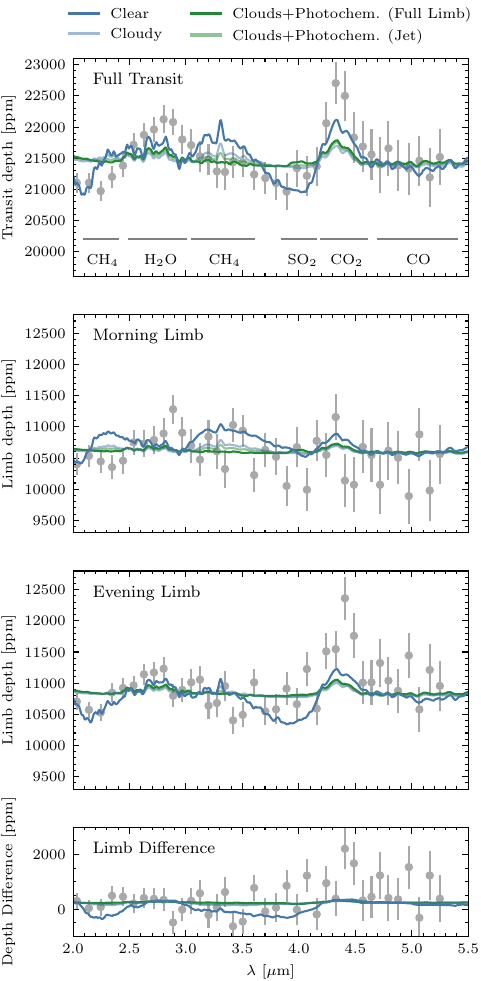}
    \caption{Transmission spectra for the ExoRad/IWF~Graz model comparing the clear and cloudy equilibrium chemistry cases with two spectra that combine clouds and photochemistry (\BOne\,and \BTwo) post-processing, with the observational data in gray. Including clouds and disequilibrium chemistry simultaneously overall improves the match to the observations. }
    \label{fig:trans_photochem+clouds}
\end{figure}
In the previous sections, we demonstrated that both clouds and disequilibrium chemistry can have a strong influence on the morning and evening spectra. Figure \ref{fig:trans_photochem+clouds} shows spectra from the ExoRad model in which the 2D VULCAN photochemistry and the IWF~Graz clouds are simultaneously applied. The cloud deck cuts off most spectral features, leaving only the peaks of the strongest spectral features like the 3.6~$\mu$m CH$_4$ feature and the 4.3~$\mu$m CO$_2$ feature visible. Introducing disequilibrium chemistry further reduces the amplitude of the CH$_4$ feature (\BTwo~post-processing) or leads to its complete disappearance (\BOne~post-processing). In the \BOne case, hints of an SO$_2$ feature at the morning limb and in the full transit spectrum appear.
Generally, the combination of clouds and disequilibrium chemistry strongly mutes any wavelength dependence in the limb difference signal.

Overall, combining clouds and disequilibrium chemistry in the ExoRad/IWF~Graz model results in a worse fit to observations at the evening terminator due to the strong suppression of spectral features compared to considering disequilibrium chemistry by itself (best match) or clouds by themselves (poor match but still better than combined). For the morning terminator, all models provide an acceptable fit (see also Table~\ref{tab:chisquared} in Appendix~\ref{sec:chisquared}). We note, however, that for a cloud model with overall less cloud opacity \citep[e.g., due to removal of iron species, ][]{Carone2026}, combining clouds and disequilibrium chemistry, which are expected to occur simultaneously, could potentially still improve the match to observations.

\section{Discussion}
\label{sec:discussion}
\subsection{Summary of Model Performance and Remaining Discrepancies}
When we compare the terminator temperatures predicted by the GCMs in this work---both their values and the differences between the terminators---we find remarkable consistency between models, even though they use significantly different assumptions. When photochemistry is applied to these atmospheres, all the models are able to roughly match the average difference between the morning and evening terminators. However, all models significantly under-predict the amplitude of the limb asymmetry in the \ce{CO2} feature at 4.3~$\mu$m, primarily due to under-predicting the strength of this feature on the evening limb.

\subsection{Potential Causes: Unexplored GCM Assumptions}
Multiple factors that were not explored in this work could lead to an increased morning--evening temperature difference, and so limb asymmetry, within GCMs. The models in this work were run based just on the previously known system properties and were not tuned in any way to match these data. However, now that properties of the planet's atmosphere have been better constrained, parameters in future GCMs could be adjusted to bring their predictions into closer alignment with the data.

For example, Welbanks et al. (in prep.) suggest the metallicity of WASP-39b lies between roughly 10 and 50x solar based on retrieval analysis of the full JWST dataset of WASP-39b \citep{CarterMayEtAl2024DataSynthesis}. Our models generally assume 10$\times$~solar, at the lower end of this range. Increasing metallicity is expected to increase horizontal temperature contrasts \citep{ShowmanEtAl2009,kataria2015,Drummond2018} and increase the \ce{CO2} abundance \citep[e.g.,][]{Moses2013}, so changes to the assumed metallicity (and/or C/O ratio) could possibly bring the strength of the predicted \ce{CO2} feature into better agreement with the observations.

Our work also did not systematically explore assumptions about numerical dissipation or (explicit) drag, both of which can shape the morning--evening temperature differences \citep[e.g.,][]{Heng2011,DobbsDixonEtAl2012,KomacekEtAl2017,Beltz2023}.
While these factors may explain part of the discrepancy between observations and models, it is unlikely that they alone can account for a five-fold difference in the asymmetry of the \ce{CO2} feature. Furthermore, an increase in just the terminator temperature differences would not explain why there is no equally strong asymmetry observed in the \ce{H2O} feature. 

\subsection{Potential Causes: Underestimated Cloud Inhomogeneities}
One possible explanation for the discrepancy between the observed and predicted limb asymmetry in the \ce{CO2} feature is that the cloudy models, in particular, may be underestimating the differences between the morning and evening terminator. 
While \citetalias{espinoza2024inhomogeneous} and the analysis of the NIRISS/SOSS WASP-39b data by \citet{fu2025} both interpret the results as indicating a change in cloud coverage between the terminators, our GCMs demonstrate the challenge of accurately capturing the cloud distributions at the terminators. Despite the ExoRad/IWF~Graz microphysics model exhibiting a lower cloud deck at the evening terminator than at the morning terminator in line with the data, it predicts by far too strong muting of spectral features through clouds at the evening terminator. Removing iron species from the model may be able to bring the model into closer agreement with the observations \citep{Carone2026}. However, while providing a much better match to the full transit spectrum and slightly increasing the morning-evening difference in the CO$_2$ feature, the iron-free models in \citet{Carone2026} still under-predict the evening amplitude of the CO$_2$ feature. 

Due to the high computational demand of the microphysics approach of the IWF~Graz model, it was performed on a single 3D temperature structure, with a single metallicity, and its results are sensitive to various model assumptions (e.g., strength of vertical mixing, nucleation, etc.). It also lacks any feedback between the clouds and the underlying temperature structure.
Although the RM-GCM model does include this feedback, its parameterization makes a number of simplifying assumptions about the cloud distribution, limiting its predictivity \citep[see discussions in ][]{Roman2019, Roman2021, Kennedy2025}. First, the particle sizes and abundances of the silicate clouds that dominate the cloud opacity, while informed by the results of more complex models, are prescribed inputs. Second, due to the prescribed cloud thickness limit and the homogeneity of the deep atmosphere where silicates first condense, the cloud distribution is forced to be uniform in longitude for planets as cool as WASP-39b \citep[though this particular limitation does not apply to hotter planets where upper-atmosphere temperatures span condensation curves, see][]{Roman2021, triantafillides2025out}. Moreover, the maximum pressure range of the clouds is chosen rather than modeled, decoupling it from explicit physical assumptions like the strength of vertical mixing. While full microphysics-GCM coupling remains computationally challenging, future work with the RM-GCM should incorporate a more complex condensation scheme, and include 3-D transport of clouds using tracers to maximize predictive power and self-consistency \citep[][]{KomacekEtAl2022,TeinturierEtAl2024,Mehta2025}. Finally, neither of the cloud models used here explore the impact of the deep-atmosphere temperature structure on cloud formation.

We further note that \citet{ArfauxLavvas2023HazeAndCloudMicrophysics} found that Na$_2$S clouds condensing on top of haze particles at the morning terminator only contribute substantial opacity in the near-infrared water bands at the morning terminator. Neither of the cloud models presented here contains Na$_2$S, as they neglect cloud-haze interactions. Na$_2$S is not expected to nucleate efficiently enough on top of TiO$_2$ particles in the relevant temperature regime to substantially contribute to the opacity \citep[e.g.,][]{GaoEtAl2020, WoitkeEtAl2020}. While the Na$_2$S clouds in \citet{ArfauxLavvas2023HazeAndCloudMicrophysics} contribute too little opacity beyond 2 $\mu$m to change the qualitative conclusions of this study, future work aiming to also match the more recent NIRISS/SOSS limb asymmetry analysis of \citet{fu2025} should also consider Na$_2$S clouds forming on top of haze particles.

This work demonstrates the intrinsically challenging problem of correctly modeling hot Jupiter clouds, with complexity from microphysical to global scales, and motivates future work further exploring these issues, as the most likely cause for the discrepancy between the models and data.

\subsection{Potential Causes: Uncertainty in Observational Analysis}
The other possibility is that we are not yet reliably measuring the spectral asymmetry, due to the intrinsically challenging nature of this new type of observation. For example, \citetalias{espinoza2024inhomogeneous} tested various treatments for stellar limb darkening as well as different approaches for deriving morning and evening spectra (fitting two half-circles with \texttt{catwoman}, fitting the half-ingress/half-egress separately with a spherical-planet transit model) in their analysis and found that while different approaches produced similar overall results, there were mild differences in the strength of the \ce{CO2} feature in both the morning and evening terminator. In particular, the limb difference in the center of the \ce{CO2} feature could vary from $\sim$1000--2000 ppm, between the different analyses. In the reanalysis by \citet{fu2025}, the strengths of the \ce{CO2} feature fell into the range of variation between these three analyses. Reducing the \ce{CO2} feature asymmetry would bring the observations closer to the predictions, but would not fully resolve the discrepancy. 

We also note that the \ce{CO2} feature is only resolved with a few data points, all of which have relatively large error bars, and so further data on this planet will be valuable. In particular, the NIRSpec G395H data also include the \ce{CO2} feature and are at higher spectral resolution \citep{AldersonEtAl2023WASP-39bG395H}. A first look at these data by \citet{TadaEtAl2025WASP-39bLimbAsymmetries} confirms large differences in the strength of the \ce{CO2} feature between morning and evening terminator, although their approach subdivides the planet more finely, making a direct comparison of the measured amplitude challenging.

\subsection{The Path Forward: New Targets, Models, and Methods}
Finally, we point out that analyzing limb asymmetries in low-resolution spectra is a new technique and the analysis of \citetalias{espinoza2024inhomogeneous}, together with \citet{MurphyEtAl2024WASP-107b}, was the first of its kind. Additional limb asymmetry analyses of more targets, and analyses using a variety of techniques and codes in addition to \texttt{catwoman} \citep[e.g., ][]{GrantWakeford2023TransmissionStrings, TadaEtAl2025WASP-39bLimbAsymmetries, chen2025asymmetry} will increase the confidence in low-resolution limb asymmetry measurements and shed more light on what is happening on WASP-39b. Thankfully, such efforts are already well on their way \citep[e.g., ][JWST Program GO 3969, PI: Espinoza; JWST Program AR 5275, PI: Kempton]{MukherjeeEtAl2025WASP-94Ab, fu2025}.

\section{Conclusions}
\label{sec:conclusions}
In this work, we analyzed the differences between morning and evening terminators, and the resulting impact on morning and evening spectra, in 3D simulations of the atmosphere of the hot Jupiter WASP-39b using five different GCMs. To examine the impact of disequilibrium chemistry, we included a 3D chemical kinetics scheme to simulate transport-induced disequilibrium thermochemistry without photochemistry in one of the GCMs (the UM), and we used the temperature structures from all five GCMs as input for the photochemistry model 2D VULCAN. We further examined the impact of clouds using two different cloud models: a parametrized equilibrium condensation model (the cloudy RM-GCM) to investigate the impact of cloud radiative feedback on temperature structure and atmospheric dynamics, and by postprocessing the pressure--temperature profiles from one of the GCMs with a microphysics model (the IWF Graz model) that simulates the growth of cloud particles through nucleation, bulk growth, evaporation and gravitational settling. Finally, we simulated the 3D distribution of photochemical hazes with SPARC/MITgcm using a passive tracer-based haze model. Our key conclusions can be summarized as:

\begin{enumerate}
    \item Even though the temperatures within the different GCM simulations vary by several hundred Kelvin within the observable atmosphere due to various model differences, as expected, the clear-atmosphere models are remarkably consistent when it comes to predicting temperature differences between morning and evening terminators. Temperature differences typically peak near 0.01~bar with a value between 200 and 300~K for models with full wavelength-dependent radiative transfer and 400~K in the model using picket-fence radiative transfer. For lower pressures, temperature differences slightly decline.
    \item Spectra derived assuming equilibrium chemistry show prominent methane absorption features at the morning terminator for several, but not all, GCMs, with substantially less methane absorption at the evening terminator. Observations, in contrast, show no detectable methane feature at the morning terminator but exhibit a strong change in the amplitude of the \ce{CO2} feature, with a much larger \ce{CO2} feature at the evening terminator by $\approx2000$~ppm. This large difference in the \ce{CO2} feature is not reproduced by any of our models, although a few models exhibit larger differences in this wavelength region than at other wavelengths ($\approx400$~ppm compared to $\approx200$~ppm).
    \item Photochemistry and atmospheric transport have a first-order effect on observed morning--evening differences, especially for the models that exhibited strong methane absorption in the morning spectrum when equilibrium chemistry was assumed. The photochemical destruction of \ce{CH4} at low pressures ($<1$~mbar) and its homogenization through transport at intermediate pressures lead to the disappearance of these features and thus much smaller morning--evening differences in the spectrum at wavelengths $<4$~$\mu$m, in good agreement with observations. Both photochemistry and atmospheric transport also have a large impact on the abundance of \ce{SO2}, which builds up on the night side from photochemical products transported there from the day side through the equatorial jet. This leads to larger abundances at the morning terminator than at the evening terminator, though the difference is not large enough to be detectable in the NIRSpec PRISM data of \citetalias{espinoza2024inhomogeneous}. Atmospheric transport and photochemistry only have a moderate impact on the abundance of \ce{CO2}, and the amplitude of the \ce{CO2} feature barely changes when these processes are included. 
    \item Radiative feedback from condensate clouds can cool the morning terminator by over a hundred Kelvin in temperature regions probed in transmission, and both terminators by even larger amounts in deeper layers. In our simulations, cloud radiative feedback also increased the temperature differences between the morning and evening terminator at $p<$0.01~bar, but reduced temperature differences for $p>0.01$~bar. Cloud microphysics models predict much larger terminator differences in cloud particle sizes, mass fraction and composition than what is captured by the parametrized equilibrium cloud model. Nevertheless, the microphysics model over-predicts cloud opacity and the terminator differences due to clouds are not large enough to explain the difference in the amplitude of the \ce{CO2} feature observed by \citetalias{espinoza2024inhomogeneous}. However, clouds can be very sensitive to temperature structure and metallicity, and we conclude that a larger exploration of the parameter space with microphysics models as well as studies combining radiative feedback and microphysics are necessary.
    \item Unlike condensate clouds, photochemical hazes have a negligible impact on limb asymmetries for wavelengths $>2 \mu$m, except when extremely high haze production rates are invoked. At shorter wavelengths, photochemical hazes lead to a slightly larger transit radius at the morning terminator.
\end{enumerate}

Overall, we conclude that the suite of five GCMs show remarkable agreement between each other when it comes to predicting the temperature difference between morning and evening terminator, and are in good agreement with the retrieved temperature difference of $177^{+65}_{-57}$~K of \citetalias{espinoza2024inhomogeneous}. When photochemistry is included, the spectral difference between morning and evening terminator also agrees well with the observed difference at wavelengths below 4 $\mu$m. However, none of our models can explain the large difference in the amplitude of the \ce{CO2} feature at 4.3 $\mu$m. 

Cloud formation restricted to the morning terminator could explain the difference in the amplitude of the \ce{CO2} feature. This could be achieved by a sufficiently hot evening terminator (due to increased atmospheric metallicity or physics not included in this study, such as drag) or by a broader exploration of assumptions in microphysics models (e.g., vertical mixing, nucleation, cold-trapping of some species). Further observations of WASP-39b and reanalysis of the existing data may also improve our understanding of the robustness of the \ce{CO2} feature and what role observational systematics play, putting models into better agreement or potentially moving it further from reach.  Similarly, the analysis of limb asymmetry in other planets---e.g., WASP-107b, WASP-94Ab---may also help better understand the parameters that drive these asymmetries, and thus help us better understand WASP-39b. Ultimately, while we have investigated here many aspects of limb asymmetry, there remain many avenues unexplored and much work to do.

\section*{Acknowledgments}
M.S. was supported by the Heising-Simons Foundation through a 51 Pegasi b fellowship. D.A.C. was supported by the Max Planck Society. This work is based on observations made with the NASA/ESA/CSA JWST. S.-M.T. was supported by NASA Exobiology grant No. 80NSSC20K1437. S.K. acknowledges funding from the European Union H2020-MSCA-ITN-2019 under grant agreement no. 860470 (CHAMELEON). D.A.L. acknowledges financial support from the School of Physics and Astronomy and the School of GeoSciences at The University of Edinburgh. D.A.L. and D.S. acknowledge financial support and use of the computational facilities of the Space Research Institute of the Austrian Academy of Sciences. M.Z.and N.J.M. acknowledge support from a UKRI Future Leaders Fellowship (grant no. MR/T040866/1). N.J.M. acknowledges support from a Science and Technology Facilities Funding Council Small Award (grant no. ST/T000082/1). J.K. acknowledges financial support from Imperial College London through an Imperial College Research Fellowship grant. B.R. received support for program number JWST-AR-05370 through a grant from the STScI under NASA contract NAS5-03127. L.G. acknowledges financial support from the Severo Ochoa grant CEX2021-001131-S funded by MCIN/AEI/10.13039/501100011033 and 
Ministerio de Ciencia e Innovación through the project PID2022-137241NB-C43.
The data were obtained from the Mikulski Archive for Space Telescopes at the Space Telescope Science Institute, which is operated by the Association of Universities for Research in Astronomy, Inc., under NASA contract NAS 5-03127 for JWST. These observations are associated with program JWST-ERS-01366. Support for program JWST-ERS-01366 was provided by NASA through a grant from the Space Telescope Science Institute.
The computational results with the ExoRad/IWF Graz model have been achieved using the Austrian Scientific Computing (ASC) infrastructure, including the Vienna Scientific Cluster (VSC). The computations using SPARC/MITgcm were performed on the HPC system Vera at the Max Planck Computing and Data
Facility (MPCDF).
We acknowledge the use of LLMs, in particular Claude Sonnet, for minor editing for clarity, grammar and sentence structure during the revisions process.

\vspace{5mm}
\facilities{JWST}

\software{astropy \citep{2013A&A...558A..33A,2018AJ....156..123A,2022ApJ...935..167A}, FastChem \citep{stock2018fastchem, stock2022fastchem},  gCMCRT \citep{lee20223d}, HELIOS \citep{malik2017helios} , matplotlib \citep{hunter2007matplotlib}, pandas \citep{mckinney2010data}, numpy \citep{2020NumPy-Array}, tqdm \citep{da2019tqdm}, aeolus \citep{sergeev_2023}, iris \citep{hattersley_2023}, VULCAN \citep{tsai2017,tsai2021} }

\section*{Data Availability}
The two-dimensional pressure-temperature and velocity profiles derived from the GCM simulations and used as input for 2D VULCAN, as well as the spectra presented in this publication, are archived at Zenodo \citep{Steinrueck_2025_Zenodo}.

\appendix

\section{Atmospheric Circulation of Clear-Atmosphere Models}\label{sec:circulation}

\begin{figure*}
    \centering
    \includegraphics{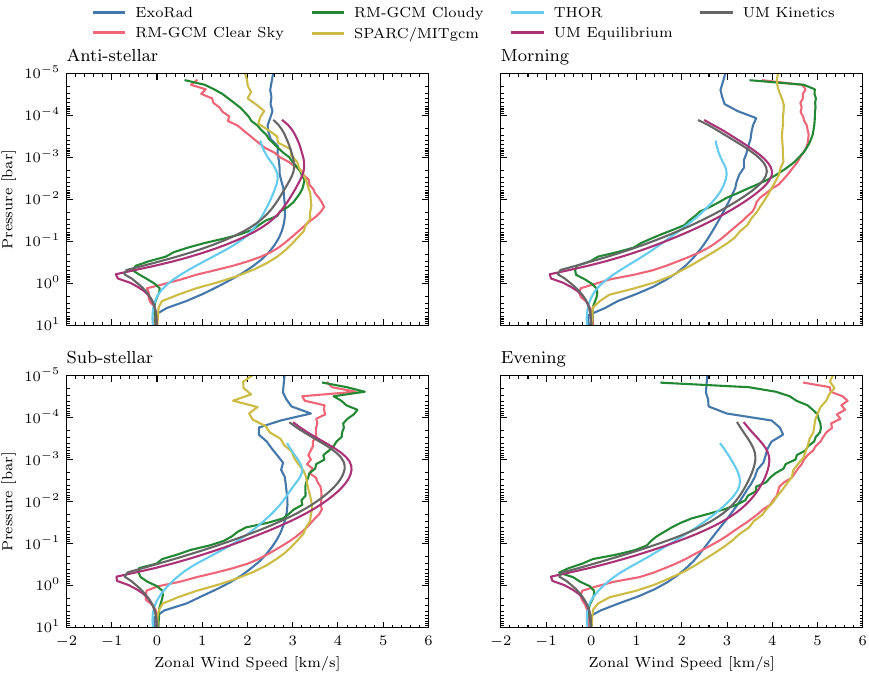}
    \caption{Vertical profiles of the zonal wind speed.  The profiles are averaged between the latitudes of $-30\degree$ and $+30\degree$.  The profiles show a 1-2~km~s$^{-1}$ variation between clear sky models at the same pressure level, with the upper atmospheres of some models showing a slowing of the jet with decreasing pressure, possibly due to boundary effects and the use of a sponge layer. }
    \label{fig:vel_profiles}
\end{figure*}

\begin{figure*}
    \centering
    \includegraphics{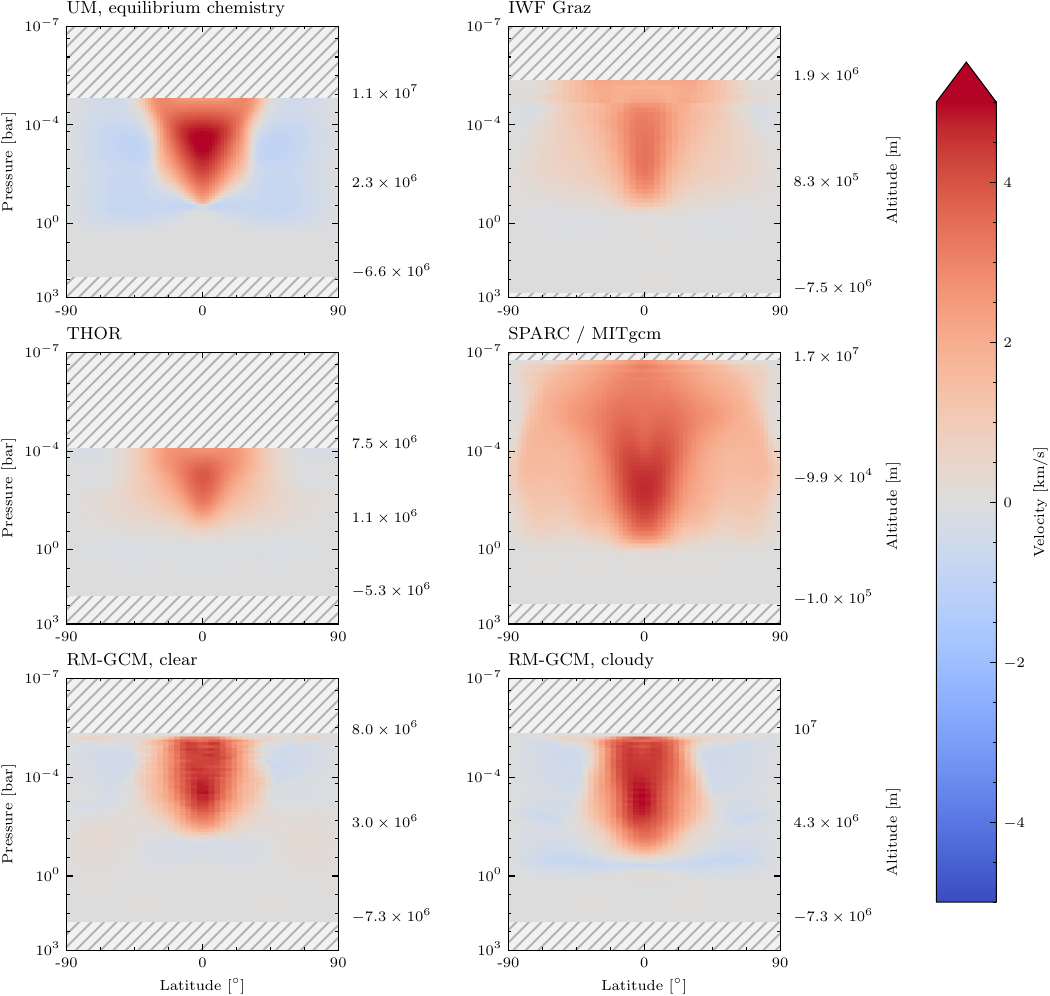}
    \caption{Mean zonal winds at the end of each simulation. Positive (red) values indicate super-rotation while negative (blue) values indicate counter-rotation.  The mean altitude is shown along the right vertical axes with an arbitrary zero point for each GCM.  }
    \label{fig:zonal_vel_profiles}
\end{figure*}

The photochemical post-processing takes the velocity profiles as input and so we provide a description of these data for completeness, with the  discussion of the photochemical results found in Section \ref{subsec:vulcanresults}.  The zonal winds, averaged between latitudes of $-30\degree$ and $+30\degree$, are shown in Figure \ref{fig:vel_profiles} at different locations within the atmosphere, and the mean zonal velocities for all latitudes are shown in Figure \ref{fig:zonal_vel_profiles}. All the simulations exhibit a super-rotating jet, ranging from 2 to 6 $\mathrm{km\,s^{-1}}$, with a deeper counter-rotating jet seen in the UM and RM-GCM simulations, due, in part, to a lack of numerical drag included at the bottom of the computational domain to maintain stability as well as model momentum exchange with the interior \citep[see ][]{LiuEtAl2013}.  The UM and, to a lesser extent THOR, also exhibit a peak in zonal windspeeds with a sharp decrease at $\sim$ 1 mbar at all longitudes considered  due to the use of a vertical sponge (see \citealt{christie2024} for a discussion for the case of the UM).  While all GCMs show some decrease in zonal windspeed within the upper few layers, this is not as extended as the decrease seen in the UM and THOR.     

\begin{figure*}
    \centering
    \includegraphics{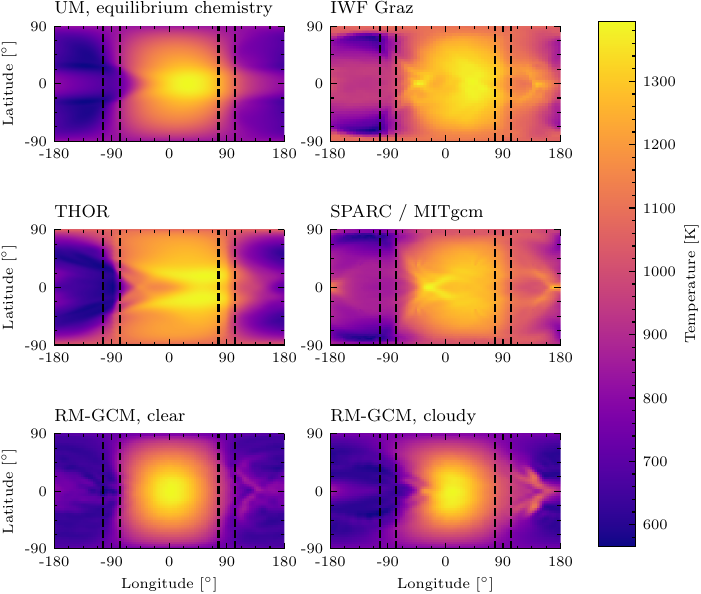}
    \caption{Gas temperatures at $10^{-3}$ bar at the end of each simulation.  The plots are oriented such that the substellar point is in the center of the panel.  The black dashed lines indicate the opening angle along the terminator, computed following \citet{Wardenier2022}.
    }
    \label{fig:temperatures_gcms}
\end{figure*}

For reference, we also provide horizontal temperature maps of the underlying GCMs in Figure \ref{fig:temperatures_gcms}.

\section{Sensitivity of photochemistry results to extending the upper boundary}
\label{sec:photochem_upper_boundary}
\begin{figure*}
   \centering
   \includegraphics[width=0.45\linewidth]{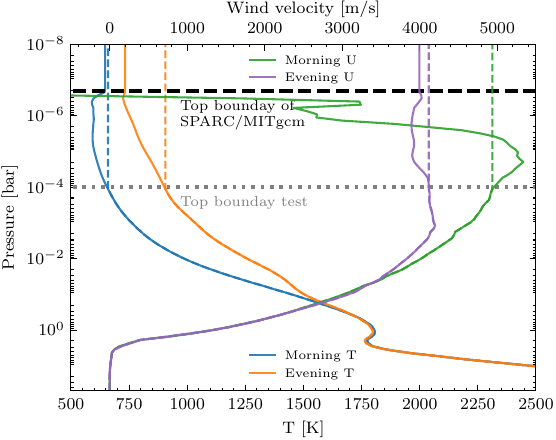}
   \includegraphics[width=0.45\linewidth]{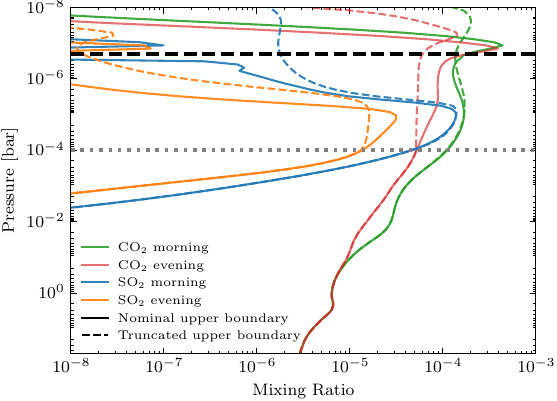}
   \caption{Left: The temperature and zonal wind profiles from SPARC/MITgcm (solid) and those held constant from 10$^{-4}$ bar (dashed) for our domain extension test run. The black dashed line indicates the level of the original top boundary of SPARC/MITgcm and the grey dotted line indicates the level of 10$^{-4}$ bar for the test run. Right: same as the left panel but for the corresponding volume mixing ratios of \ce{CO2} and \ce{SO2}.}
   \label{fig:cut_01mbar}
\end{figure*}
The major limitation in modeling photochemistry using GCM inputs is that their upper boundaries generally do not extend to sufficiently low pressures, constrained by numerical stability requirements. The GCMs employed in our study had upper boundaries ranging from 10$^{-4}$ bar to 2$\times$10$^{-7}$ bar, with THOR using the lowest-altitude top boundary and SPARC/MITgcm using the highest. To properly cover the optically thin region in the UV, our 2D photochemical model uses a top boundary at 10$^{-8}$ bar. When adopting the temperatures and winds from the GCMs into VULCAN 2D, we simply extrapolated these properties by holding them constant from the GCMs' upper boundaries. Here, we test the impacts of extending the 2D photochemical model domain beyond the GCMs' limits. Using SPARC/MITgcm as a control model, we performed an additional 2D VULCAN simulation where we held temperatures and winds constant from the pressure level corresponding to THOR's upper boundary.

Figure \ref{fig:cut_01mbar} shows that the extension of the upper boundary can lead to minor errors, especially for \ce{SO2}, which is sensitive to both local temperature and atmospheric circulation. Nevertheless, the discrepancy is mostly confined to the extrapolated region above 0.1 mbar and does not propagate downward. This test suggests that the model results are reliable up to the upper boundary of the 3D GCM. For SPARC/MITgcm, the top boundary at 2$\times$10$^{-7}$ bar lies above the region that is probed by the transmission spectroscopy.

\section{Detailed definitions of cloud properties}
\label{sec:cloud_properties}
In Table \ref{tab:cloud_definitions}, we summarize the differences between the two cloud models used in this work, the ExoRad/IWF~Graz model and the cloudy RM-GCM.

\begin{deluxetable*}{cccc}
\tablecaption{Definition and units of quantities that are used in cloud models discussed in this work. }
\label{tab:cloud_definitions}
\tablehead{\colhead{Variable} & \colhead{Meaning or Definition} & \colhead{ExoRad/IWF~Graz}  & \colhead{RM-GCM} \\  
& & \colhead{(Sect.~\ref{ssubsec:IWF_clouds})} &  \colhead{(Sect. ~\ref{subsec:RMGCM_cloud_description})} }
\startdata
$T_{\rm gas}$ & Local gas temperature &  Depends on gas opacity only  & Depends on gas and cloud opacity\\
$p_{\rm gas}$ & Local gas pressure &   &  \\
$\epsilon_{\rm i}$ & Gas elemental abundance
& 10$\times$ Solar${}^\mathrm{a}$ & 10$\times$ Solar${}^\mathrm{b}$\\
$a$ & Particle size & & \\
$f(a)$  & Particle size distribution &   N/A & Monodisperse \\
$L_j$ & Dust moments &   $\rho_{\rm gas} L_j=\int V^{j/3}f(V)dV$ & \\
$\langle a\rangle$                   & Mean particle size & $\langle a\rangle = \sqrt[3]{{\frac{3}{4\pi}}} \frac{L_1(p_{\rm gas})}{L_0(p_{\rm gas})}$ & $0.1\mu$m for $p_{\rm gas}<10^{-2}$ bar, \\
& $\displaystyle=\frac{\int_a f(a)\, a \,da}{\int_a f(a)\,da}$ &  & $80\mu$m  for $p_{\rm gas} > 80$ bar\\
& &  &  $\left<a\right>$ varies linearly with $p_\mathrm{gas}$ in between\\ 
&   &\\
$n_{\rm d}$ & Cloud particle number density &    $= \rho_{\rm  gas}L_0$ &  from total cloud mass, $\rho_{\rm d}$, and $a$\\
$V_{\rm s}$ & Material volume of kind s &  Eqs. 2, 9 in \cite{Helling2008_Paper1_new} & Eq.2 in \cite{Roman2019} \\
$s$   & Cloud material species & 16 & 8  \\
& (e.g. Fe[s], Mg$_2$SiO$_4$[s], $\ldots$) & &  \\
$\rho_{\rm d}$                            & Cloud particle mass density &   &  \\
& $=\sum_{\rm s} n_{\rm d,s} M_{\rm s}$ for $a=a_0=$const  & \\
& $=\sum_{\rm s}\int_V f(V)\,\rho_{\rm s}V_{\rm s}(V)dV$  & \\
$M_{\rm s}$ & Mass of particle of material $s$ &  & Eq.1 in \cite{Roman2021} \\
\enddata
\tablenotetext{a}{\citet{Asplund2009}}
\tablenotetext{b}{\citet{Anders1989}. We note that both sets of solar element abundances differ in the oxygen abundance.}

\end{deluxetable*}%

\section{Differences between cloud-post processing here and in Espinoza et al. (2024)}
\label{sec:spec_differences}

\begin{figure}
    \centering
    \includegraphics{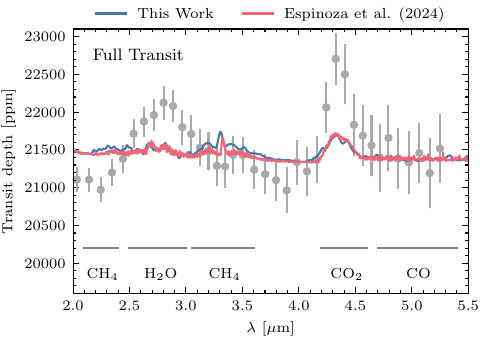}
    \caption{Comparison of the post-processed ExoRad/IWF-Graz spectra used in this work (see Section~\ref{subsec:spectra_description}), shown in blue, and in \citetalias{espinoza2024inhomogeneous} shown in red.}
    \label{fig:IWF_post-processing_comp}
\end{figure}

To uniformly post-process the GCMs in this work, the method of generating spectra from the ExoRad/IWF-Graz cloud models has been altered relative to that used in \citetalias{espinoza2024inhomogeneous}. As shown in Figure~\ref{fig:IWF_post-processing_comp}, although the spectra calculated from the two methods tend to agree overall, there exist minor deviations between the two. In the 2--5 micron range, these differences are chromatic and are on the order of 100 ppm. This value compares favorably to the median uncertainty of the transmission spectrum in this wavelength range (270 ppm for the full transit spectrum at $R=100$).

A number of differences in the calculation of these two model spectra could explain their mild disagreement. Perhaps the most obvious of these differences lies in geometry. The spectra shown in \citetalias{espinoza2024inhomogeneous} were generated with \texttt{petitRADTRANS} \citep{Molliere2019}, which is by nature a 1D radiative transfer code. That is, it assumes that atmospheric quantities (temperature, chemistry, and clouds) vary only as a function of the radial pressure coordinate. In contrast, the 3D calculations presented in this work account for heterogeneity in the thermal and chemical structure of the input GCM, along with changes in cloud properties across multiple longitudes (see Sec.~\ref{subsec:IWF_cloud_description}). The variation in these quantities along the transit chord can nonlinearly affect the strength of spectral lines relative to the continuum and to one another. Such ``3D effects'' are well known in the exoplanets literature \citep[e.g.,][]{caldas2019,MacDonaldLewis2022Trident,nixon2022aura}, and it is reasonable to assume that they are at play in the spectrum of WASP-39~b.  

Other details of the radiative transfer may account for the rest of the deviations between these two spectra. For instance, the two post-processing approaches draw from different spectral line lists (Table~\ref{tab:opacitydata}), which may impart subtle changes in spectral feature strength. Furthermore, the \texttt{petitRADTRANS} calculations were performed with the correlated-k approximation with a resolution of $R=1000$, whereas the 3D code used opacity sampling before convolving with a smoothing kernel and binning to instrumental resolution. Other differences between the radiative transfer codes \citep[such as altitude gridding, quoted precision of universal constants, and local calculation of gravity within cells; e.g.,][]{barstow2020comparison} may also contribute. However, a detailed investigation of these factors is beyond the scope of our work.

\section{Line Lists}
\label{sec:opacitydata}

The GCM simulations using correlated-k radiative transfer --- ExoRad, SPARC/MITgcm, and the Unified Model --- rely on opacity data from a variety of sources, as does the post-processing code of \citet{kempton2012constraining}.  The line lists for the opacity sources used in the Unified Model are found in \citet{Goyal2020} while for SPARC/MITgcm the opacities come from \citet{FreedmanEtAl2008,FreedmanEtAl2014} and \citet{LupuEtAl2014}, as described in Section \ref{sec:sparc}. For reference, we collect the line lists used by ExoRad and the post-processing code in Table \ref{tab:opacitydata}.

\begin{deluxetable*}{lcc}
\tablecaption{Line Lists}
\label{tab:opacitydata}
\tablenum{}
\tablehead{\colhead{Source} & \colhead{ExoRad} & \colhead{Post-processing}  } 
\startdata
\multicolumn{3}{l}{\em Atomic and Molecular Sources} \\
\ce{H2O} &  \citet{Polyansky2018}  & \citet{Polyansky2018} \\
\ce{CH4} & \citet{yurchenko17} & \citet{yurchenko17} \\
\ce{CO2} & \citet{yurchenko2020exomol}  & \citet{yurchenko2020exomol}\\
\ce{NH3} & \citet{Coles_2019_NH3} & \citet{Coles_2019_NH3} \\
\ce{CO} & \citet{li2015} & \citet{li2015} \\
\ce{H2S} & \citet{azzam2016} & - \\
\ce{HCN} & \citet{Barber2014} & - \\
\ce{PH3} & \citet{Sousa_Silva_2014} &  -\\
\ce{FeH} & \citet{Wende2010} & - \\
\ce{C2H2} & - & \citet{chubb2020exomol} \\
\ce{SO2} & - & \citet{underwood2016exomol} \\
\ce{Na} &  \citet{Piskunov1995} & - \\
        & \citet{allard1995}  &  \\
\ce{K} & \citet{Piskunov1995} & -\\
       & \citet{allard1995}  & \\
\\
\multicolumn{3}{l}{\em Rayleigh Scattering} \\
\ce{H2} & \citet{Dalgarno1962} & \citet{haynes2016crc}  \\
\ce{He} &  \citet{Chan1965} & \citet{haynes2016crc} \\
\ce{N2} & - & \citet{haynes2016crc}  \\
\ce{H2O} &  - & \citet{haynes2016crc} \\
\ce{CO} & - & \citet{haynes2016crc}  \\
\ce{CO2} &  - & \citet{haynes2016crc} \\
\ce{NH3} &  - & \citet{haynes2016crc} \\
\ce{CH4} &  - & \citet{haynes2016crc} \\
\ce{O2} &  - & \citet{haynes2016crc} \\
\ce{O3} &  - & \citet{haynes2016crc} \\
\\
\multicolumn{3}{l}{\em Collision-Induced Absorption} \\
\ce{H2}--\ce{H2} &\citet{Richard2012} & \citet{Richard2012} \\
\ce{H2}--\ce{He} & \citet{Richard2012}& \citet{Richard2012} \\
\enddata

\end{deluxetable*}

\section{Reduced Chi-Squared}
\label{sec:chisquared}
We collect here in Table \ref{tab:chisquared} for reference the reduced chi-squared metrics ($\chi_\mathrm{red}^2$) for each synthetic spectrum generated relative to the observations.

\begin{deluxetable*}{lccc}
\tablecaption{The $\chi_\mathrm{red}^2$ of synthetic spectra relative to observations} 
\label{tab:chisquared}
\tablehead{\colhead{Simulation} & \colhead{Full Transit} & \colhead{Morning Limb}  & \colhead{Evening Limb}}
\startdata
\multicolumn{4}{l}{\em Clear Sky} \\
ExoRad & 1.33 & 1.52 & 1.77\\
RM-GCM Clear Sky & 0.68 & 1.16 & 1.34\\
SPARC/MITgcm & 2.52 & 1.16 & 1.34\\
THOR & 0.63 & 1.00 &  1.50\\
UM Equilibrium &  1.25 &  1.16 & 1.34 \\
UM Kinetics&  1.09 & 1.11 & 1.77 \\
\\
\multicolumn{4}{l}{\em Cloudy}\\
ExoRad/IWF Graz & 2.06 & 1.14 & 1.99\\
RM-GCM Cloudy &0.83 &1.16 &1.35\\
\\
\multicolumn{4}{l}{\em \BOne \,Photochemistry}\\
ExoRad & 0.82 & 1.12 & 1.20\\
RM-GCM Clear & 0.83 & 1.20& 1.17\\
SPARC/MITgcm & 0.94& 1.37 & 1.42\\
THOR & 0.64 & 1.04& 1.43\\
UM Kinetics & 2.13 & 1.75 & 1.46\\
\\
\multicolumn{4}{l}{\em \BTwo\, Photochemistry}\\
ExoRad & 0.42 & 1.20& 1.41\\
RM-GCM Clear & 0.32 & 0.92&  1.20\\
SPARC/MITgcm & 0.41& 1.55& 2.11\\
THOR & 0.91& 1.02 & 1.70\\
UM Kinetics & 1.05& 1.21& 1.41\\
\\
\multicolumn{4}{l}{\em Photochemistry + Clouds}\\
ExoRad/IWF Graz + \BOne\, Photochem. & 2.33 & 1.16 & 2.26 \\
ExoRad/IWF Graz + \BTwo\, Photochem. & 2.38 & 1.15 & 2.27 \\
\enddata
\end{deluxetable*}%

\bibliography{biblio_limbasymmetries}
\bibliographystyle{aasjournal}

\end{document}